\documentclass[twocolumn]{aastex631}
\usepackage{rotating}
\usepackage{xcolor, soul}

\newcommand {\kms}{$\rm{km\ s^{-1}}$}

\begin{document}

\title{Ejecta, Rings, and Dust in SN 1987A with JWST MIRI/MRS}

\correspondingauthor{O. C. Jones}
\email{olivia.jones@stfc.ac.uk}

\author[0000-0003-4870-5547]{O.\ C.\ Jones}
\affiliation{UK Astronomy Technology Centre, Royal Observatory, Blackford Hill, Edinburgh, EH9 3HJ, UK}

\author[0000-0001-6872-2358]{P.\ J.\ Kavanagh}
\affil{Department of Experimental Physics, Maynooth University, Maynooth, Co. Kildare, Ireland}
\affil{Dublin Institute for Advanced Studies, School of Cosmic Physics, Astronomy \& Astrophysics Section 31 Fitzwilliam Place, Dublin 2, Ireland}

\author[0000-0002-3875-1171]{M.\ J.\ Barlow}
\affiliation{Department of Physics and Astronomy, University College London (UCL), Gower Street, London WC1E 6BT, UK}

\author[0000-0001-7380-3144]{T.\ Temim}
\affil{Department of Astrophysical Sciences, Princeton University, Princeton, NJ 08544, USA}

\author[0000-0001-8532-3594]{C.\ Fransson}
\affiliation{Department of Astronomy, Stockholm University, The Oskar Klein Centre, AlbaNova, SE-106 91 Stockholm, Sweden}

\author[0000-0003-0065-2933]{J.\ Larsson}
\affiliation{Department of Physics, KTH Royal Institute of Technology, The Oskar Klein Centre, AlbaNova, SE-106 91 Stockholm, Sweden}

\author[0000-0002-5797-2439]{J.\ A.\ D.\ L.\ Blommaert}
\affiliation{Astronomy and Astrophysics Research Group, Department of Physics and Astrophysics, Vrije Universiteit Brussel, Pleinlaan 2, B-1050 Brussels, Belgium}

\author[0000-0002-0522-3743]{M.\ Meixner}
\affil{Jet Propulsion Laboratory, California Institute of Technology, 4800 Oak Grove Dr., Pasadena, CA 91109, USA }

\author[0000-0003-0778-0321]{R.\ M.\ Lau}
\affil{NSF’s NOIR Lab 950 N. Cherry Avenue, Tucson, AZ 85721, USA}

\author[0000-0001-9855-8261]{B.\ Sargent}
\affiliation{Space Telescope Science Institute, 3700 San Martin Drive, Baltimore, MD 21218, USA}
\affiliation{Center for Astrophysical Sciences, The William H. Miller III Department of Physics and Astronomy, Johns Hopkins University, Baltimore, MD 21218, USA}

\author{P.\ Bouchet}
\affiliation{Laboratoire AIM Paris-Saclay, CNRS, Universit\'e Paris Diderot, F-91191 Gif-sur-Yvette, France}
\affil{Université Paris-Saclay, Université Paris Cité, CEA, CNRS, AIM, F-91191 Gif-sur-Yvette, France }

\author[0000-0002-4571-2306]{J.\ Hjorth}
\affil{DARK, Niels Bohr Institute, University of Copenhagen, Jagtvej 128, 2200 Copenhagen, Denmark}

\author[0000-0001-7416-7936]{G.\ S.\ Wright}
\affiliation{UK Astronomy Technology Centre, Royal Observatory, Blackford Hill, Edinburgh, EH9 3HJ, UK}

\author[0000-0001-6492-7719]{A.\ Coulais}
\affiliation{LERMA, Observatoire de Paris, Universit\'e PSL, Sorbonne Universit\'e, CNRS, Paris, France}
\affil{Université Paris-Saclay, Université Paris Cité, CEA, CNRS, AIM, F-91191 Gif-sur-Yvette, France }

\author[0000-0003-2238-1572]{O.\ D.\ Fox}
\affiliation{Space Telescope Science Institute, 3700 San Martin Drive, Baltimore, MD, 21218, USA}

\author{R.\ Gastaud}
\affiliation{Université Paris-Saclay, CEA, DEDIP, 91191, Gif-sur-Yvette, France}

\author[0000-0002-2041-2462]{A.\ Glasse}
\affiliation{UK Astronomy Technology Centre, Royal Observatory, Blackford Hill, Edinburgh, EH9 3HJ, UK}

\author[0000-0002-2667-1676]{N.\ Habel}
\affil{Jet Propulsion Laboratory, California Institute of Technology, 4800 Oak Grove Dr., Pasadena, CA 91109, USA}

\author[0000-0002-2954-8622]{A.\ S.\ Hirschauer}
\affil{Space Telescope Science Institute, 3700 San Martin Drive, Baltimore, MD 21218, USA}

\author[0000-0002-0577-1950]{J.\ Jaspers}
\affil{Dublin Institute for Advanced Studies, School of Cosmic Physics, Astronomy \& Astrophysics Section 31 Fitzwilliam Place, Dublin 2, Ireland}
\affil{Department of Experimental Physics, Maynooth University, Maynooth, Co. Kildare, Ireland}

\author{O.\ Krause}
\affiliation{Max-Planck-Institut fuer Astronomie, Koenigstuhl 17, D-69117 Heidelberg, Germany}

\author[0000-0003-4023-8657]{L.\ Lenki\'{c}}
\affil{Stratospheric Observatory for Infrared Astronomy, NASA Ames Research Center, Mail Stop 204-14, Moffett Field, CA 94035, USA}

\author[0000-0001-6576-6339]{O.\ Nayak}
\affil{Space Telescope Science Institute, 3700 San Martin Drive, Baltimore, MD 21218, USA}

\author[0000-0002-4410-5387]{A.\ Rest}
\affil{Space Telescope Science Institute, 3700 San Martin Drive, Baltimore, MD 21218, USA}
\affil{Department of Physics and Astronomy, Johns Hopkins University, 3400 North Charles Street, Baltimore, MD 21218, USA}

\author{T.\ Tikkanen}
\affiliation{School of Physics \& Astronomy, Space Research Centre, University of Leicester, Space Park Leicester, 92 Corporation Road, Leicester LE4 5SP, UK}

\author[0000-0002-4000-4394]{R.\ Wesson}
\affiliation{School of Physics and Astronomy, Cardiff University, Queen’s Buildings, The Parade, Cardiff, CF24 3AA, UK}


\author[0000-0002-9090-4227]{L.\ Colina}
\affil{Centro de Astrobiolog\'{\i}a (CAB), CSIC-INTA, Ctra. de Ajalvir km 4, Torrej\'on de Ardoz, E-28850, Madrid, Spain}

\author[0000-0001-7591-1907]{E.\ F.\ van Dishoeck}
\affil{Max-Planck Institut f\"ur Extraterrestrische Physik (MPE),Giessenbachstr. 1, D-85748, Garching, Germany}
\affil{Leiden Observatory, Leiden University, 2300 RA Leiden, The Netherlands}

\author[0000-0001-9818-0588]{M.\ G\"udel}
\affil{Dept. of Astrophysics, University of Vienna, T\"urkenschanzstr. 17, A-1180 Vienna, Austria}
\affil{Max-Planck-Institut für Astronomie (MPIA),K\"onigstuhl 17, D-69117 Heidelberg, Germany}
\affil{ETH Z\"urich, Institute for Particle Physics and Astrophysics, Wolfgang-Pauli-Str. 27, 8093 Z\"urich, Switzerland}

\author[0000-0002-1493-300X]{Th.\ Henning}
\affil{Max-Planck-Institut für Astronomie (MPIA),K\"onigstuhl 17, D-69117 Heidelberg, Germany}

\author{P.-O.\ Lagage}
\affiliation{Laboratoire AIM Paris-Saclay, CNRS, Universit\'e Paris Diderot, F-91191 Gif-sur-Yvette, France}
\affil{Université Paris-Saclay, Université Paris Cité, CEA, CNRS, AIM, F-91191 Gif-sur-Yvette, France }

\author[0000-0002-3005-1349]{G.\ \"Ostlin}
\affil{Department of Astronomy, Stockholm University, The Oskar Klein Centre, AlbaNova, SE-106 91 Stockholm, Sweden}

\author[0000-0002-2110-1068]{T.\ P.\ Ray}
\affil{Dublin Institute for Advanced Studies, School of Cosmic Physics, Astronomy \& Astrophysics Section 31 Fitzwilliam Place, Dublin 2, Ireland}

\author[0000-0002-1368-3109]{B.\ Vandenbussche}
\affil{Institute of Astronomy, KU Leuven, Celestijnenlaan 200D, 3001 Leuven, Belgium}


\begin{abstract}
Supernova (SN) 1987A is the nearest supernova in $\sim$400 years. Using the {\em JWST} MIRI Medium Resolution Spectrograph, we spatially resolved the ejecta, equatorial ring (ER) and outer rings in the mid-infrared 12,927~days (35.4~years) after the explosion.
The spectra are rich in line and dust continuum emission, both in the ejecta and the ring. 
Broad emission lines (280-380~km~s$^{-1}$ FWHM) seen from all singly-ionized species originate from the expanding ER, with properties consistent with dense post-shock cooling gas. 
Narrower emission lines (100-170~km~s$^{-1}$ FWHM) are seen from species originating from a more extended lower-density component whose high ionization may have been produced by shocks progressing through the ER, or by the UV radiation pulse associated with the original supernova event.
The asymmetric east-west dust emission in the ER has continued to fade, with constant temperature, signifying a reduction in dust mass. Small grains in the ER are preferentially destroyed, with larger grains from the progenitor surviving the transition from SN into SNR. 
The ER dust is fit with a single set of optical constants, eliminating the need for a secondary featureless hot dust component. 
We find several broad ejecta emission lines from [Ne~{\sc ii}], [Ar~{\sc ii}], [Fe~{\sc ii}], and [Ni~{\sc ii}]. With the exception of [Fe~{\sc ii}]~25.99$\mu$m, these all originate from the ejecta close to the ring and are likely being excited by X-rays from the interaction. The [Fe~{\sc ii}]~5.34$\mu$m to 25.99$\mu$m line ratio indicates a temperature of only a few hundred K in the inner core, consistent with being powered by ${}^{44}$Ti decay.
\end{abstract}




\section{Introduction} 
\label{sec:intro}

Supernovae (SNe) are important contributors to the chemical evolution of the universe over cosmic time, however, their relative net contribution to dust production in the early universe is still debated. 
While extant studies of SNe have shown that these events are prodigious sources for dust production \citep[e.g.][]{Matsuura2011, Niculescu-Duvaz2022, Shahbandeh2023}, they have found that their ability to also destroy dust is similarly (and potentially equally) proficient \citep[e.g.,][]{Gall2018, Slavin2020, Kirchschlager2023}.
Assessing the effects of SNe on their host galaxies, and understanding how their dust production and destruction processes occurs over months to years, requires time-series observations at a high angular resolution to build realistic models.

SN 1987A provides a once-in-a-lifetime opportunity to investigate the evolution of a supernova into a supernova remnant \citep[SNR;][]{McCray2016}. 
On February 23, 1987, light from a massive blue supergiant (BSG) Sanduleak -69 202 in the Large Magellanic Cloud (LMC), $\sim$50 kpc from us \citep{Pietrzynski2019}, which collapsed and exploded as a supernova was visible to the naked eye \citep[e.g.,][]{Arnett1989}; producing an initial burst of neutrinos consistent with the formation of a neutron star \citep[][]{Burrows1988}. 
In the intervening years, SN 1987A has been extensively studied across the electromagnetic spectrum revealing important insights into the physics of supernovae, the properties of stellar explosions, and the evolution of stars. See Table~\ref{tab:87aevents} for a listing of some key events in the evolution of SN~1987A.

The SN 1987A morphology is comprised of asymmetrical inner ejecta, a dense equatorial ring (ER) approximately 2 arcsec in diameter inclined at 43$^{\circ}$ from the line-of-sight, and two faint outer rings 5 arcsec in diameter.
The outer rings, thought to be produced at the same time as the ER, have an expansion velocity of 26 km~s$^{-1}$ \citep{Crotts2000}, whilst the ejecta is expanding at a rate of $\sim$ 2000--10000 km s$^{-1}$ \citep{McCray1993, Fransson2013, Larsson2016, Kangas2022}. 

Emission from the ER is the dominant component of SN 1987A observations from the X-ray to the mid-infrared (IR). The ER was initially photoionized by the SN shock breakout, but that emission has faded away and the emission from the ER is now dominated by shock interaction.
The ER was likely produced in a binary merger event during the progenitor's red supergiant phase of evolution \citep{Morris2009} and was initially expanding at 10.3 km s$^{-1}$ relative to the SN. As shocks from the blast wave interacted with the dense gas in the ER, $\sim$30 hot spots began to form and steadily brighten in the ring in optical wavelengths \citep{Pun1995, Bouchet2000, Fransson2015, McCray2016}, increasing the expansion velocity of the ring to $\sim$700 km s$^{-1}$ \citep{Larsson2019a, Kangas2023}. Once the initial shock from the blast wave passed (at $\sim$7,500 days, or 20.5 yrs) these hot spots began to fade and there is increasing interaction outside the ER as the forward shock expands outwards. Interactions with reverse shocks propagating into the SN ejecta also affect recent processing, gradually destroying the ring in the process \citep{Fransson2015,Larsson2019a, Orlando2019, Arendt2020, Kangas2022}.


Using the T-ReCS instrument, \cite{Bouchet2004} resolved 10 $\mu$m emission from SN 1987A at day 6067 (year 16.6) tracing its warm dust morphology; this emission is dominated by the ER, with less than 10$^{-5}$ M$_{\odot}$ of warm progenitor dust.
{\em Spitzer} InfraRed Spectrograph (IRS) observations revealed the ER is comprised of warm ($T_d \approx$ 180 K) silicate dust \citep{Bouchet2006, Dwek2010, Arendt2016}. This emission brightened up to day 8708 (year 23.8) as the dust grains were collisionally heated by the shocked gas, however, the shape of the dust spectrum remained unchanged.   

Cryogenic {\em Spitzer} IRAC fluxes (obtained from day 6130, year 16.8) appeared to be strengthening in agreement with the IRS spectra, however, a secondary hot $T_d  > 350$ K dust component of undetermined composition appeared to be necessary to fit the blue-end (3 -- 8 $\mu$m) of the spectral energy distribution (SED; \citealt{Dwek2010}). 
Imaging with the warm {\em Spitzer} mission continued to monitor the SN 1987A light curves until day 11,885 (year 32.5) \citep{Arendt2020}. These appear to peak between day 8500--9000 (year 23.2-24.6) before undergoing a gradual fading in luminosity since day 9810 (year 26.9) post-explosion, which is consistent with the passage of the forward shock through the ER and the gradual destruction of dust emitting at these shorter wavelengths.

In the IR, CO and SiO vibration bands from the ejecta became apparent in the integrated spectra of SN 1987A from day 200 post-explosion \citep{Roche1991,Spyromilio1988}, and dust formation in the ejecta occurred between 400 and 500 days, consequently obscuring some of the optical and near-IR emission and re-radiating this light in the mid-IR \citep{Lucy1989, Bouchet1991, Wooden1993}. This dust was characterised by a featureless single-temperature graybody emission spectrum.

The {\em Herschel} far-IR observations \citep{Matsuura2011} unveiled a large reservoir of newly condensed cold dust with temperature T$_{\rm d} = 17-23$ K and mass M$_{\rm d} = 0.3-0.8$ M$_{\odot}$, depending on whether its composition arises solely from amorphous carbon or a mix of amorphous carbon and silicates  \citep{Matsuura2015}. 
The origin of this dust was unambiguously confirmed to be from the inner ejecta of SN 1987A by the Atacama Large Millimetre Array (ALMA; \citealt{Indebetouw2014, Matsuura2017}). This extremely efficient dust condensation in the ejecta (measured 24 years after the explosion) requires almost all the expected Mg and Si produced in the SN to condense into grains and is significantly higher than the dust mass measured by \citet{Bouchet1991} and \citet{Wooden1993} in the first few years post-explosion. 
This ejecta dust, 36 years after the explosion, has not yet experienced processing by the reverse shock \citep{France2010, Chevalier2017}; it remains unclear how much of this dust will survive and dissipate into the interstellar medium once this interaction occurs.

The SN 1987A ejecta has a complex asymmetric morphology. In the optical, the ejecta is elongated north-south with a faint central region \citep{Larsson2011}, and is currently undergoing a free expansion interior to the ER \citep{Larsson2016, Larsson2019b}.  ALMA observations show the CO and SiO molecules in the ejecta have different distributions, with cold dust emission coincident with the faint optical region \citep{Kamenetzky2013, Indebetouw2014}. These observations agree with theoretical models of dust formation in SN that show CO and SiO form in different regions of the ejecta a few hundred days after explosion \citep{Sarangi2013, Sluder2018}.

Several near- and mid-IR lines from the ejecta were observed during the first years after explosion with both ground-based instruments and the Kuiper Airborne Observatory. These have proved to be extremely useful as diagnostics of the temperature, density and abundances in the ejecta \citep{Aitken1988,Rank1988,Spyromilio1990,Wooden1993,Colgan1994}. In particular, the line ratios of the [\ion{Fe}{2}] lines indicated a low filling factor for the iron in the core of the ejecta \citep{Li1993,Kozma1998}, indicative of ${}^{56}$Ni heating, during the first weeks after explosion, creating low-density bubbles in the core. Also the [\ion{Fe}{2}] line profiles were very useful for probing the asymmmetry and abundance mixing of the ejecta \citep{Spyromilio1990,Haas1990}. These observations have provided some of the most important constraints on the explosion mechanism. 

The launch of {\em JWST} \citep{Rigby2023, Gardner2023} with its superb sensitivity and high spatial resolution images and integral field spectrographs (IFS) has opened a new era in mid-IR observations, enabling us to obtain IR data on SN 1987A at a comparable resolution to the optical {\em HST} data and sub-mm ALMA data. For instance, NIRSpec observations reveal asymmetry in the ejecta, likely due to a binary merger prior to explosion and morphological 3D structures in the circumstellar medium  \citep{Larsson2023}.

With the Mid-Infrared Instrument (MIRI; \citealt{Wright2023}) Medium Resolution Spectrometer (MRS) onboard {\em JWST}, we can finally observe and spatially resolve the ER and ejecta of SN 1987A at mid-IR wavelengths with a high S/N ratio. 
The paper is organised as follows. In Section~\ref{sec:obs} we describe the observations and data reduction. 
Our results are presented in Section~\ref{sec:results}.
Finally, in Section~\ref{sec:discussion} we discuss the implications of these data.  
The conclusions are summarised in Section \ref{sec:conclusions}. 

\begin{table*}
\caption{Evolution of SN 1987A. See text and reviews by \cite{McCray1993} and \cite{Fransson2016} for further details and references. } 
\label{tab:87aevents}
\centering
\begin{tabular}{lll}
\hline
\hline
Year & Epoch & Event/observations \\
&  (days) & \\
 \hline
 1987 & 0 &  Explosion and flash ionization of the rings.  \\
 1987 & 200 & Near-IR detection of CO and SiO vibrational emission from the ejecta. \\
1988 & 450 & Mid-IR detection of dust formation in the ejecta. \\
1990 & 1200 & Blast wave starts interacting with \ion{H}{2} region inside the ER, radio and X-ray emission detected.  \\
1997 & 3700 & Blast wave hits the ER -- first optical hotspot (forward shock) and reverse shock detected. \\
\nodata & \nodata & Hotspots appear all around the ER and the ER brightens at all wavelengths.   \\
2003 & 6070 & Mid-IR detection of spatially resolved dust emission from the ER. \\
2010 & 8500 & Detection of cold dust far-IR/submm emission by {\em Herschel}; ejecta origin confirmed by ALMA in 2012. \\
2010 & 8300 & The optical emission from the ER starts fading, the blast wave has passed the ER.  \\
\nodata & \nodata & The ER fades at most wavelengths. \\
2013 & 9500 &  First detection of optical spots outside the ER.\\
\hline
\end{tabular}
\end{table*}

\section{Observations and data reduction}
\label{sec:obs}

SN 1987A was observed with the MIRI/MRS on 2022-07-16 as part of the guaranteed time programme \#1232 (PI: Wright).  The observations consist of 94 groups of three integrations using the {\sc fastr1} readout pattern for all three bands (SHORT, MEDIUM, LONG), which provided complete spectral coverage from 4.9 to 27.9 $\mu$m at medium resolution (R $\sim$  4000--1500; \citealt{Jones2023}). As no wavelength channel was prioritised for the SN 1987A MRS observations, a 4-point extended source dither pattern in the `negative' direction optimised for `ALL' channels was selected. The MRS fields of view range from 3.2$\arcsec$ $\times$ 3.7$\arcsec$ in channel 1 to 6.6$\arcsec$ $\times$ 7.7$\arcsec$ in channel 4 \citep{Wells2015, Argyriou2023arXiv}. 
No target acquisition was employed and no dedicated background observations were taken. The total MRS integration time corresponded to 9396s, or approximately 3.75 hr including all overheads. 

Simultaneous MIRI imaging in the F560W, F770W, and F1000W filters was obtained to improve the astrometric accuracy of the MRS data.  These observations were taken using the {\sc slowr1} readout pattern (to reduce data volume), with 10 groups and 3 integrations, using the same dither pattern as the MRS data. 

\begin{figure*}[!ht]
\centering
\includegraphics[width=\hsize]{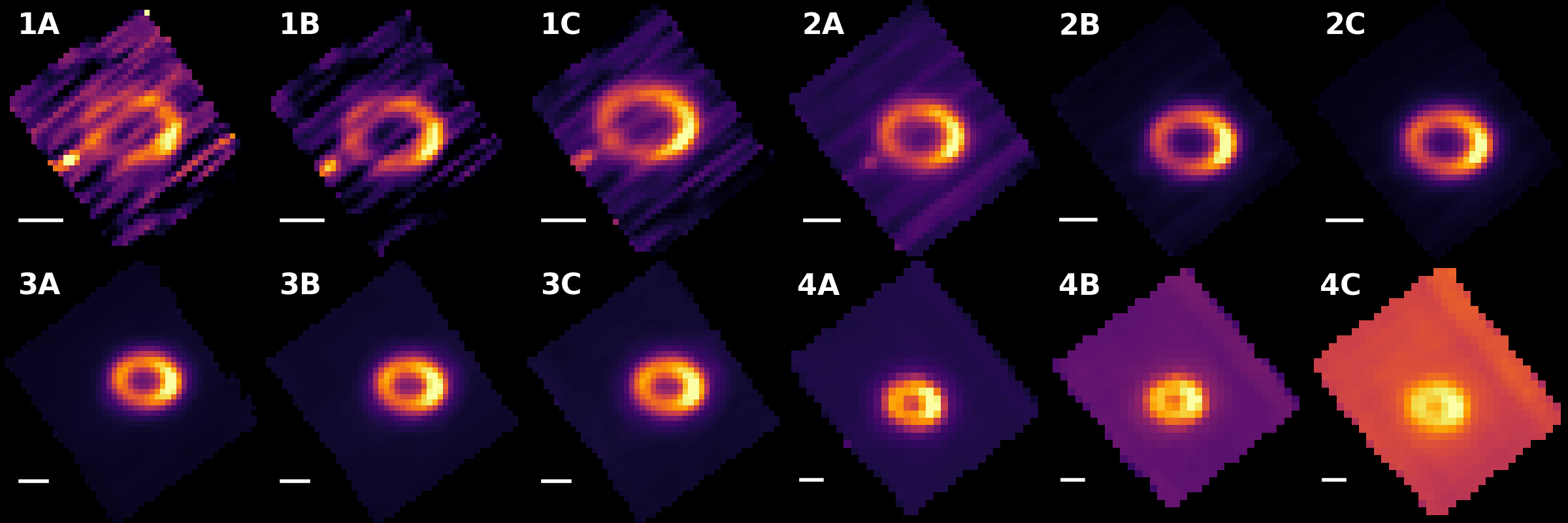}
\caption{Sample slices from our 12 MRS sub-band cubes, with the band labels shown in the top-left. The white lines in each pane represent 1.0~arcsec to highlight the increasing size of the FOV, as well as the decreasing spatial resolution from channels 1 to 4. `Star 3' is visible to the lower left of the ER in bands 1A--2A. North is up, east is left.} 
\label{Fig:cube_det_feat}
\end{figure*}

To reduce our MRS data we used a development version of the JWST Calibration Pipeline \citep{2023zndo...7692609B} retrieved on May-23-2023\footnote{\url{https://github.com/spacetelescope/jwst}}, with versions 11.17.0 and `jwst\_1090.pmap' of the Calibration Reference Data System (CRDS) and CRDS context, respectively, and development versions of the FLT-5 MRS calibration files which include an improved photometric calibration derived from the Cycle 1 calibration program. We ran all raw data files through the \texttt{Detector1Pipeline} to convert the ramp files into rate images. Since inaccuracies in the spacecraft pointing information can be introduced by guide star catalogue errors and roll uncertainty \citep[see][]{Pont2022}, we corrected the world coordinate system (WCS) reference keywords in our rate images by determining the offset of point sources detected in the F560W simultaneous imaging field to their Gaia Data Release 3\footnote{\url{https://www.cosmos.esa.int/web/gaia/dr3}} counterparts. This resulted in an offset correction of $-2.825\arcsec$ in RA and $-0.630\arcsec$ in Dec. This correction is expected to be accurate to $\lesssim$0.1$\arcsec$ \citep{Patapis2023}, with the upper limit due to uncertainties in the relative astrometry between the MRS and imager. The pointing correction must be applied before running stage 2 of the pipeline as the WCS solution is attached to the files in its first step.  

We ran all files through \texttt{Spec2Pipeline}, with the optional \texttt{residual\_fringe} step switched on to produce flux calibrated rate images, from which we constructed spectral cubes for all of the 12 MRS sub-bands using \texttt{Spec3Pipeline}. MRS data usually contain large numbers of `warm' pixels which are not flagged in the bad pixel mask as they evolve over time. For data where no dedicated background is available, these can have a significant impact on cubes and extracted spectra, where they manifest as spurious emission lines. To account for these when producing our cubes, we used the updated \texttt{outlier\_detection} step in \texttt{Spec3Pipeline} to capture and omit these pixels from the cube-building process. We show sample slices from each of the 12 sub-band cubes in Fig.~\ref{Fig:cube_det_feat} to highlight some features of our MRS data. 

Channel 1 and 2 cubes suffer from a `striping' in the x-dimension of the cube frame, which is residual dark current. Even though the pipeline subtracts a dark current reference from the data, drifts in the effective dark current over time cause such residuals. Unfortunately, without dedicated background observations, this residual cannot be removed without risking the subtraction of real signal from SN1987A. The channel 4 cubes suffer from an increased signal towards the right edge in the cube frame. This likely results from a more uncertain flux calibration near the slice edges. We omitted these regions from our analysis. 

\begin{figure*}
\centering
\includegraphics[width=\hsize]{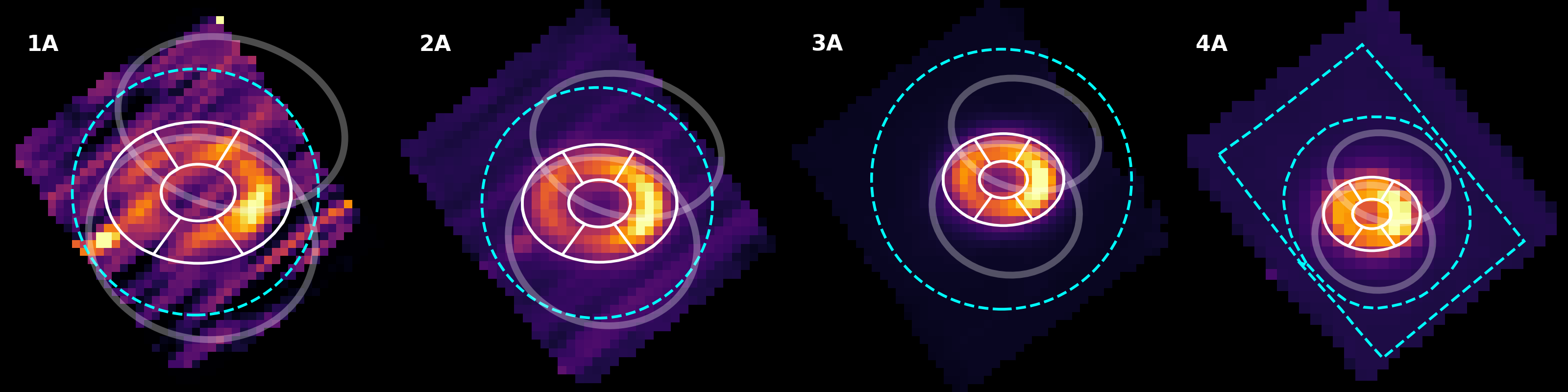}
\includegraphics[width=\hsize]{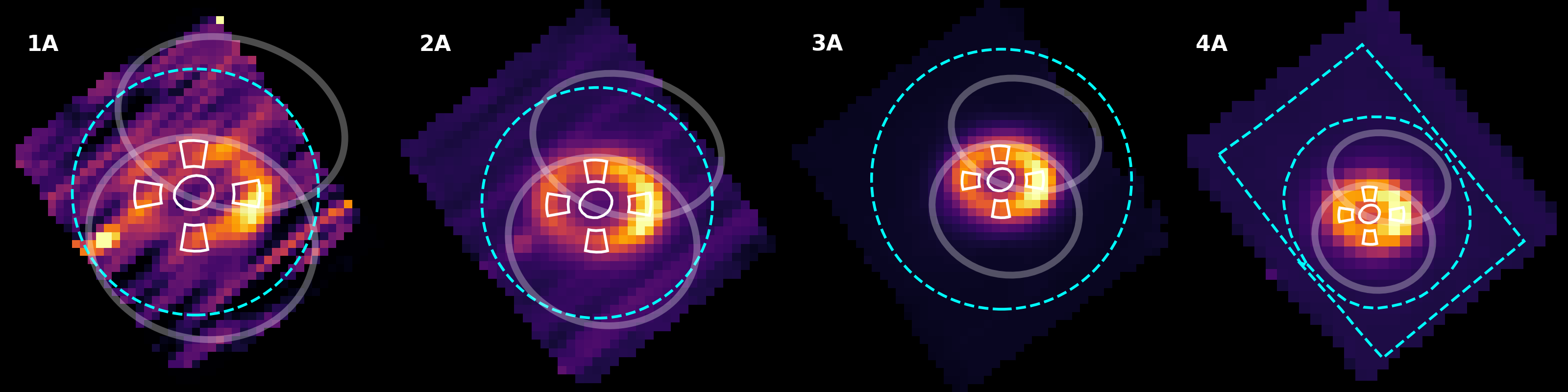}
\caption{Top row: The spectra extraction regions for the ER spectra plotted on band A in each channel. The total ER spectrum was extracted from the white elliptical annulus, with the east and west ER extracted from the segments on the left and right. Bottom row: Same as top but for the cardinal point and ejecta spectra. In both rows, the location of the outer rings is shown by the two faded ellipses, the channels 1--3 background inner boundaries are shown by the dashed cyan circles, and the channel 4 background region is shown by the dashed cyan polygon. The `entire' SN~1987A system spectrum was extracted from a region slightly smaller than the inner boundary of the background regions and excluding Star~3. North is up, east is left.} 
\label{Fig:spec_extraction_Regions}
\end{figure*}

Two final aspects of our data to contend with were the degradation in spatial resolution and increasing background from band 1A through 4C, both of which are evident in Fig.~\ref{Fig:cube_det_feat}. The former presents challenges in creating representative full-band MRS spectra from small regions of the SN~1987A system. These are discussed in more detail below. As we had no dedicated background pointing we were required to use regions within the sub-band FOVs to create a background that could be subtracted from our spectra. We defined circular regions surrounding SN~1987A for channels 1--3, with the backgrounds extracted from all pixels exterior to the circle. For channel 4 we used a polygonal region surrounding SN~1987A but omitting the bright edge of the cubes mentioned previously. These background regions are shown in Fig.~\ref{Fig:spec_extraction_Regions}, along with the location of the outer rings defined using their morphology in the [Ne~{\sc ii}]~12.81354~$\mu$m line. It is clear from Fig.~\ref{Fig:spec_extraction_Regions} that the outer rings take up a significant fraction of the channel 1 and 2 FOVs making it impractical to have a `true' background. The sizes of background boundaries in these channels are therefore a trade-off between producing a representative background while mitigating contamination from the line emission from outer rings (see Sect.~\ref{sec:lines}). 

Due to the complex morphology of SN~1987A, we have extracted spectra from several locations in the MRS cube. These include 1) the total ER; 2) segments of equal size from the east and west sides of the ER; 3) four smaller cardinal point regions from the N, S, E and W portions of the ER, comparable in size to the hot-spots observed in the optical, 4); the central ejecta, and finally 5); the entire SN~1987A system to enable comparisons with past {\em Spitzer} observations\footnote{The {\em Spitzer} IRS spectra can be downloaded from \url{https://irsa.ipac.caltech.edu/applications/Spitzer/SHA/}}. All regions are shown in Fig.~\ref{Fig:spec_extraction_Regions}. Spectral regions were defined using SAO~Image DS9 \citep{Joye2003} and extracted using the \texttt{aperture} module in the Astropy Photutils package\footnote{\url{https://photutils.readthedocs.io/en/stable/}} \citep{Bradley2022} by parsing the regions using the Astropy Regions package\footnote{\url{https://astropy-regions.readthedocs.io/en/stable/}} \citep{Bradley2022b}. We applied an additional post-pipeline residual fringe correction to our spectra to account for the high-frequency fringing in channels 3 and 4, thought to originate in the MIRI/MRS dichroics, which are not effectively removed in the pipeline. This correction is now included in the JWST Calibration Pipeline package under the \texttt{extract\_1d} step. Finally, to produce the total, full-band MRS spectrum of SN~1987A we scaled the MRS sub-band fluxes for minor offsets by determining the median flux difference in the overlapping wavelength regions, starting from the band 1A. These sub-band offsets were typically at the $~\sim1$\% level, with the worst-case offset of $~\sim8$\% being at the band 2C/3A overlap region where MRS fringing is most problematic. The scaled sub-band spectra were stitched using the \texttt{combine\_1d} tool in the JWST Calibration Pipeline package.

\section{Results}
\label{sec:results}

With the MRS we have detected and spatially resolved the 5--28 $\mu$m emission from the ejecta, ER and outer rings of SN 1987A at day 12,927 (year 35.4) after the explosion. The ER is well resolved and isolated from the nearby Stars 2 (which is not visible in the mid-IR) and 3 (Figure~\ref{Fig:cube_det_feat}) which have affected lower spatial resolution observations of the system, especially at shorter wavelengths. 
The ER produces most of the IR emission, however, this emission has a complex morphology and is not evenly distributed over the ring. It is brightest on the west side. This is consistent with the evolution of the peak hot spot emission observed by {\em HST} in the ER from the northeast to the west side circa 2006 \citep{Fransson2016, Larsson2019a}, due to the asymmetric distribution of matter in the ER.  Similar behaviours are also observed in the X-ray \citep{Frank2016} and near-IR \citep{Kangas2023,Larsson2023}.

\begin{figure*}
\centering
    \includegraphics[width=\hsize,angle=0]{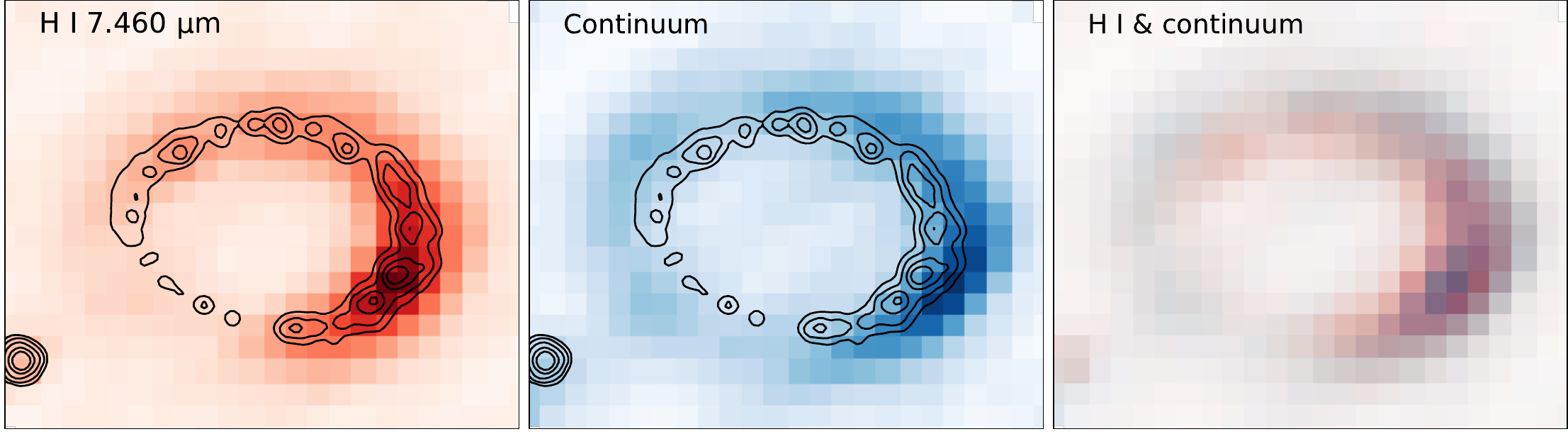}
    \caption{Illustration of the different spatial distributions of the line emission and continuum. The left panel shows an integrated image of the \ion{H}{1} 7.460\ $\mu$m line after subtraction of the continuum, the middle panel shows an image of the continuum on each side of the line, and the right panel show the two images together. The black contours in the left and middle panels show the hotspots in a HST/F657N image obtained on 2022-09-06 (dominated by H$\alpha$ emission, Rosu et al.~in preparation). Note the good agreement between the \ion{H}{1} 7.460\ $\mu$m line and the  {\em HST} image, while the continuum is more spatially extended.  
    }
    \label{Fig:line_vs_cont}
\end{figure*}

The line emission from the ER originates from dense clumps of gas, observed as hotspots in the optical images, while we find that the continuum emission is more spatially extended. This is illustrated for the \ion{H}{1}~7.460~$\mu$m Pfund-$\alpha$ line and nearby continuum in Figure~\ref{Fig:line_vs_cont}. In order to quantify the differences in the spatial distribution, we fit the ER with an elliptical model with a Gaussian radial profile and sinusoidal intensity in the azimuthal direction. While this model does not account for all of the substructure in the ER, it captures the bulk of the emission and hence provides a good estimate of the overall size.  The semi-major (minor) axes of the best-fit ellipses are $0\farcs{850} \pm 0\farcs{006}$ ($0\farcs{632} \pm 0\farcs{002}$) for the line image, compared to $0\farcs{988} \pm 0\farcs{004}$ ($0\farcs{721} \pm 0\farcs{002}$) for the continuum image.

In the collapsed MIRI/MRS data cubes the overall morphology is comparable to the T-ReCS 10~$\mu$m data observed by \cite{Bouchet2004, Bouchet2006} at days 7241, 7565, 8720 and VLT/VISIR 10--18~$\mu$m images at days 10950--10976 (year 30.0) \citep{Matsuura2022}.
Higher spatial resolution {\em JWST} mid-IR images of SN 1987A obtained on the same day post-explosion as the MIRI/MRS data including the ER morphology are discussed in depth in \citet{Bouchet2024}. Strikingly the resolved emission from the [Ne {\sc ii}] lines reveals the outer ring structure (see Sect.~\ref{sec:lines}) and the central compact ejecta is seen in the [Ar {\sc ii}] and  [Fe {\sc ii}] emission lines (see Sect.~\ref{sec:ejecta}).

\begin{figure*}
\centering
\includegraphics[width=\hsize]{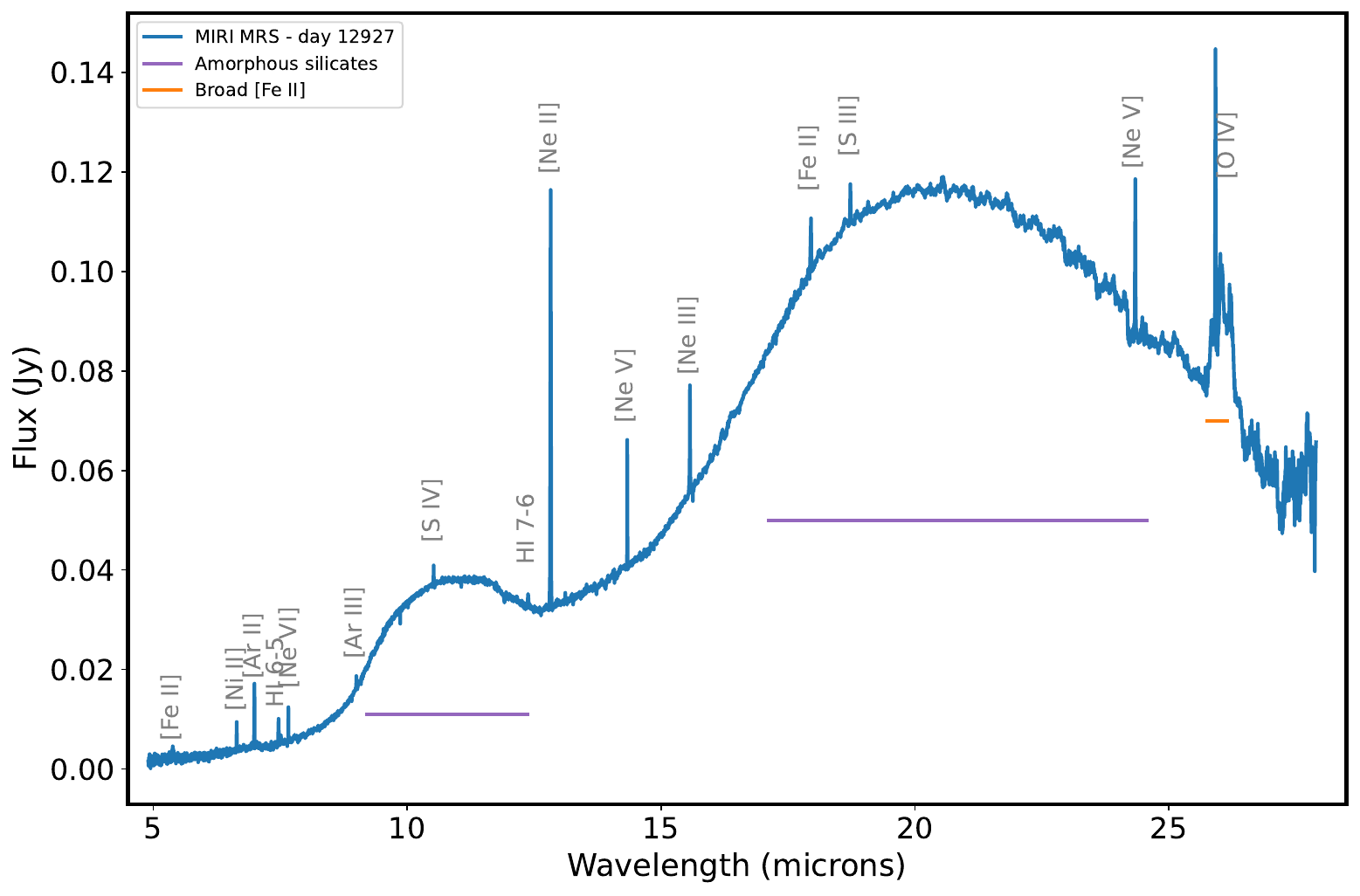}
\caption{The total MIRI MRS background subtracted spectrum of SN 1987A 12,927 days (35.4 years) after the explosion. The mid-IR spectrum at this epoch is dominated by dust emission from amorphous silicates in the inner ER. 
The line emission is from a combination of the ionised ER and the central ejecta (with a broad emission profile, e.g., [Fe~{\sc ii}]  at 26~$\mu$m). The most significant lines are highlighted.} 
\label{Fig:spec}
\end{figure*}

Figure~\ref{Fig:spec} shows the total background-subtracted MRS spectrum of SN~1987A, for the region defined by the white annulus shown in the upper part of Fig.~\ref{Fig:spec_extraction_Regions}. The spectra are rich in atomic emission lines and silicate dust. These spectral features and their interpretation are discussed in detail in the following sections.

\begin{table*}
\caption{Detected emission lines, their measured heliocentric radial velocities, full-width half maxima, and line fluxes, for the whole ER extraction (subtending a solid angle of $1.072200\times10^{-10}$ sr). } 
\label{tab:whole_ring}
\centering
\begin{tabular}{ccccc}
\hline
\hline
Species & $\lambda_{\rm lab}$\footnote{Vacuum wavelengths from the compilation by \cite{vanHoof2018}} & Velocity & FWHM & Flux \\ 
 & ($\mu$m) & (km~s$^{-1}$) & (km~s$^{-1}$) &
 ($10^{-24}$~W~cm$^{-2}$) \\
 \hline
{[}Ni~{\sc ii}{]} & 6.6360 & 266.4$\pm$2.4 & 303.7$\pm$5.6 & 232.3$\pm$3.8 \\
{[}Ar~{\sc ii}{]} & 6.985274 & 262.5$\pm$1.7 & 288.8$\pm$4.1 & 483.0$\pm$6.0 \\
{[}Na~{\sc iii}{]} & 7.31775 & 274.6$\pm$22.3 & 354.7$\pm$44.6 & 30.7$\pm$3.9 \\
H~{\sc i}+He~{\sc i} 6-5\footnote{H~{\sc i} 8-6 7.502493~$\mu$m was also detected} & 7.459858 & 244.6$\pm$5.4 & 344.1$\pm$12.4 & 207.8$\pm$6.9 \\
{[}Ne~{\sc vi}{]} & 7.6524 & 277.2$\pm$1.7 & 106.6$\pm$4.4 & 85.6$\pm$2.9 \\
{[}Ar~{\sc iii}{]} & 8.99138 & 258.9$\pm$5.7 & 166.4$\pm$15.4 & 35.7$\pm$2.7 \\ 
{[}S~{\sc iv}{]} & 10.51049 & 286.8$\pm$2.4 & 105.5$\pm$5.6 & 33.6$\pm$1.6 \\
{[}Ni~{\sc ii}{]} & 10.6822 & 229.5$\pm$14.1 & 285.2$\pm$43.8 & 20.9$\pm$2.4 \\
H~{\sc i}+He~{\sc i} 7-6 & 12.371898 & 238.9$\pm$8.7 & 384.4$\pm$23.4 & 66.1$\pm$3.5 \\
{[}Ne~{\sc ii}{]} & 12.813548 & 275.3$\pm$1.3 & 297.0$\pm$3.3 & 1425.0$\pm$14.0 \\
{[}Ne~{\sc v}{]} & 14.32168 & 273.1$\pm$0.7 & 98.9$\pm$1.7 & 94.5$\pm$1.4 \\
{[}Ne~{\sc iii}{]} & 15.5551 & 275.2$\pm$2.5 & 170.9$\pm$7.7 & 142.1$\pm$4.9 \\
{[}Fe~{\sc ii}{]} & 17.936026 & 242.8$\pm$5.9 & 280.2$\pm$15.8 & 117.8$\pm$5.6 \\
{[}S~{\sc iii}{]} & 18.71303 & 254.6$\pm$9.6 & 145.0$\pm$14.5 & 33.5$\pm$3.9 \\
{[}Ne~{\sc v}{]} & 24.3175 & 278.7$\pm$3.7 & 137.6$\pm$9.7 & 92.6$\pm$5.4 \\
{[}O~{\sc iv}{]} & 25.8903 & 342.7$\pm$5.9 & 122.6$\pm$12.4 & 124.9$\pm$12.2 \\

\hline
\end{tabular}
\end{table*}

\begin{sidewaystable*}
\vspace{-9cm}
\setlength\tabcolsep{2pt}
\caption{Detected emission lines in the four cardinal point spectra, their measured heliocentric radial velocities, full-width half maxima, and line fluxes. } 
\label{tab:cardinal_points}
\centering
\begin{tabular}{@{}cccccccccccccc@{}}
\hline
\hline
 & & \multicolumn{3}{c}{Ring-East ($3.616917\times10^{-12}$ ster)} & \multicolumn{3}{c}{Ring-North ($3.614933\times10^{-12}$ ster)} & \multicolumn{3}{c}{Ring-West
  ($3.617161\times10^{-12}$ ster)} & \multicolumn{3}{c}{Ring-South
  ($3.614884\times10^{-12}$ ster)} \\
 Species & $\lambda_{\rm lab}$\footnote{Vacuum wavelengths from the compilation by \cite{vanHoof2018}}
 & Velocity & FWHM & Flux
 & Velocity & FWHM & Flux
 & Velocity & FWHM & Flux
 & Velocity & FWHM & Flux
 \\ 
 & ($\mu$m) & (km~s$^{-1}$) & (km~s$^{-1}$) &
 {\footnotesize $10^{-24}$~W/cm$^2$}
 & (km~s$^{-1}$) & (km~s$^{-1}$) &
 {\footnotesize $10^{-24}$~W/cm$^2$}
 & (km~s$^{-1}$) & (km~s$^{-1}$) &
 {\footnotesize $10^{-24}$~W/cm$^2$}
 & (km~s$^{-1}$) & (km~s$^{-1}$) &
 {\footnotesize $10^{-24}$~W/cm$^2$}
\\
 \hline
  & & & & & & & & & & & & & \\
{[}Ni~{\sc ii}{]} & 6.6360 & 335.3$\pm$5.2 & 225.4$\pm$12.2
& 5.35$\pm$0.26 
& 144.0$\pm$1.9 & 212.0$\pm$4.6 & 14.2$\pm$0.26
& 250.0$\pm$0.9 & 230.7$\pm$2.1 & 37.2$\pm$0.29
& 429.1$\pm$3.2 & 212.5$\pm$7.0 & 7.49$\pm$0.23 \\
 & & & & & & & & & & & & & \\
{[}Ar~{\sc ii}{]} & 6.985274 & 320.0$\pm$4.1 & 239.0$\pm$10.9 & 9.22$\pm$0.34 
& 135.7$\pm$1.5 & 233.0$\pm$3.6 & 29.3$\pm$0.40
& 263.0$\pm$0.5 & 228.3$\pm$1.3 & 82.7$\pm$0.40 
& 420.1$\pm$4.0 & 215.5$\pm$9.9 & 11.9$\pm$0.47
\\
 & & & & & & & & & & & & & \\
{[}Na~{\sc iii}{]} & 7.31775 & {\bf ---} & {\bf ---} & {\bf ---} & {\bf ---} & {\bf ---} & {\bf ---} & 272.6$\pm$9.4 & 297.4$\pm$21.5 & 4.15$\pm$0.27 & {\bf ---} & {\bf ---} & {\bf ---} \\
 & & & & & & & & & & & & & \\
H~{\sc i} 6-5\footnote{H~{\sc i} 8-6 7.502493~$\mu$m was also detected in each spectrum} & 7.459858 & 307.8$\pm$7.4 & 252.9$\pm$18.1 & 3.88$\pm$0.24
& 95.0$\pm$3.6 & 258.7$\pm$8.7 & 13.5$\pm$0.40 
& 242.3$\pm$3.0 & 274.2$\pm$7.4 & 33.9$\pm$0.80
& 427.4$\pm$6.4 & 282.4$\pm$16.0 & 6.10$\pm$0.30
\\
 & & & & & & & & & & & & & \\
{[}Ne~{\sc vi}{]} & 7.6524 & 267.2$\pm$4.5 & 114.5$\pm$9.8 & 2.21$\pm$0.18
& 269.9$\pm$3.9 & 110.1$\pm$9.8 & 2.70$\pm$0.20
& 273.4$\pm$2.3 & 127.6$\pm$6.0 & 5.10$\pm$0.20
& 288.9$\pm$9.1 & 90.5$\pm$24.0 & 0.830$\pm$0.178
\\
 & & & & & & & & & & & & & \\
{[}Ar~{\sc iii}{]} & 8.99138 & 241.4$\pm$13.1 & 175.7$\pm$35.7 & 0.914$\pm$0.147
& 98.0$\pm$32.4 & 381.5$\pm$54.2 & 1.60$\pm$0.25
& 255.5$\pm$6.6 & 253.3$\pm$20.1 & 4.22$\pm$0.26
& {\bf ---} & {\bf ---} & {\bf ---}
\\ 
 & & & & & & & & & & & & & \\
{[}S~{\sc iv}{]} & 10.51049 & 288.3$\pm$7.4 & 99.4$\pm$17.6 & 0.678$\pm$0.105
& 279.0$\pm$6.2 & 101.5$\pm$11.8 & 0.876$\pm$0.104
& 297.8$\pm$4.2 & 83.9$\pm$10.6 & 1.51$\pm$0.16
& {\bf ---} & {\bf ---} & {\bf ---}
\\
 & & & & & & & & & & & & & \\
{[}Ni~{\sc ii}{]} & 10.6822 & {\bf ---} & {\bf ---} & {\bf ---} & {\bf ---} & {\bf ---} & {\bf ---} & 258.3$\pm$4.1 & 232.0$\pm$10.3 & 3.83$\pm$0.14 & {\bf ---} & {\bf ---} & {\bf ---} \\ 
 & & & & & & & & & & & & & \\
H~{\sc i} 7-6 & 12.371898 
& {\bf ---} & {\bf ---} & {\bf ---}
& 96.4$\pm$5.4 & 252.8$\pm$13.0 & 2.97$\pm$0.14 
& 256.2$\pm$3.8 & 296.4$\pm$10.1 & 8.68$\pm$0.25
& 443.5$\pm$19.9 & 501.0$\pm$63.6 & 3.21$\pm$0.33
\\
 & & & & & & & & & & & & & \\
{[}Ne~{\sc ii}{]} & 12.813548 & 299.2$\pm$1.9 & 296.1$\pm$5.1 & 30.2$\pm$0.45 
& 144.4$\pm$0.85 & 246.0$\pm$2.1 & 63.0$\pm$0.47
& 276.4$\pm$0.7 & 237.7$\pm$1.7 & 185.8$\pm$1.2
& 406.8$\pm$1.0 & 256.0$\pm$2.5 & 41.1$\pm$0.35
\\
 & & & & & & & & & & & & & \\
{[}Ne~{\sc v}{]} & 14.32168 & 269.5$\pm$1.6 & 106.6$\pm$4.1 & 2.84$\pm$0.093 
& 269.2$\pm$1.2 & 93.2$\pm$3.0 & 2.78$\pm$0.074
& 271.5$\pm$1.1 & 96.0$\pm$2.9 & 4.64$\pm$0.11
& 282.8$\pm$2.3 & 86.9$\pm$7.6 & 1.35$\pm$0.083
\\
 & & & & & & & & & & & & & \\
{[}Ne~{\sc iii}{]} & 15.5551 & 267.1$\pm$5.2 & 181.6$\pm$17.9 & 3.13$\pm$0.22 
& 231.8$\pm$6.4 & 241.1$\pm$17.5 & 6.19$\pm$0.34
& 284.4$\pm$1.7 & 187.0$\pm$4.9 & 15.7$\pm$0.33
& 375.1$\pm$4.3 & 268.4$\pm$10.9 & 3.50$\pm$0.12
\\
 & & & & & & & & & & & & & \\
{[}Fe~{\sc ii}{]} & 17.936026 & 247.4$\pm$11.4 & 275.8$\pm$31.3 & 2.67$\pm$0.25 
& 127.8$\pm$4.9 & 247.7$\pm$10.6 & 4.97$\pm$0.20
& 243.3$\pm$2.4 & 230.4$\pm$6.9 & 11.7$\pm$0.28
& 336.4$\pm$12.5 & 225.7$\pm$27.8 & 3.02$\pm$0.34
\\
 & & & & & & & & & & & & & \\
{[}S~{\sc iii}{]} & 18.71303 & 242.7$\pm$9.4 & 138.7$\pm$16.8 & 1.41$\pm$0.17
& 241.8$\pm$9.8 & 146.4$\pm$18.0 & 1.42$\pm$0.17
& 264.0$\pm$14.7 & 162.4$\pm$26.9 & 2.63$\pm$0.45 
& {\bf ---} & {\bf ---} & {\bf ---}
\\
 & & & & & & & & & & & & & \\
{[}Ne~{\sc v}{]} & 24.3175 & 276.7$\pm$5.2 & 142.2$\pm$13.0 & 3.05$\pm$0.24 
& 269.8$\pm$6.6 & 125.5$\pm$19.7 & 2.48$\pm$0.29
& 273.6$\pm$5.8 & 126.2$\pm$16.9 & 4.03$\pm$0.41
& 292.1$\pm$11.2 & 125.0$\pm$19.9 & 1.30$\pm$0.22
\\
 & & & & & & & & & & & & & \\
{[}O~{\sc iv}{]} & 25.8903 & 347.3$\pm$6.4 & 136.1$\pm$16.1 & 4.12$\pm$0.43 
& 335.3$\pm$17.1 & 143.3$\pm$40.4 & 3.70$\pm$0.95
& 337.0$\pm$6.5 & 116.4$\pm$15.2 & 5.97$\pm$0.69
& 365.7$\pm$16.4 & 198.3$\pm$58.3 & 4.09$\pm$0.89
\\
 & & & & & & & & & & & & & \\
\hline
\end{tabular}
\end{sidewaystable*}


\begin{figure*}
\centering
\includegraphics[width=\hsize,trim={2.5cm 1.5cm 1.5cm 3.cm},clip]{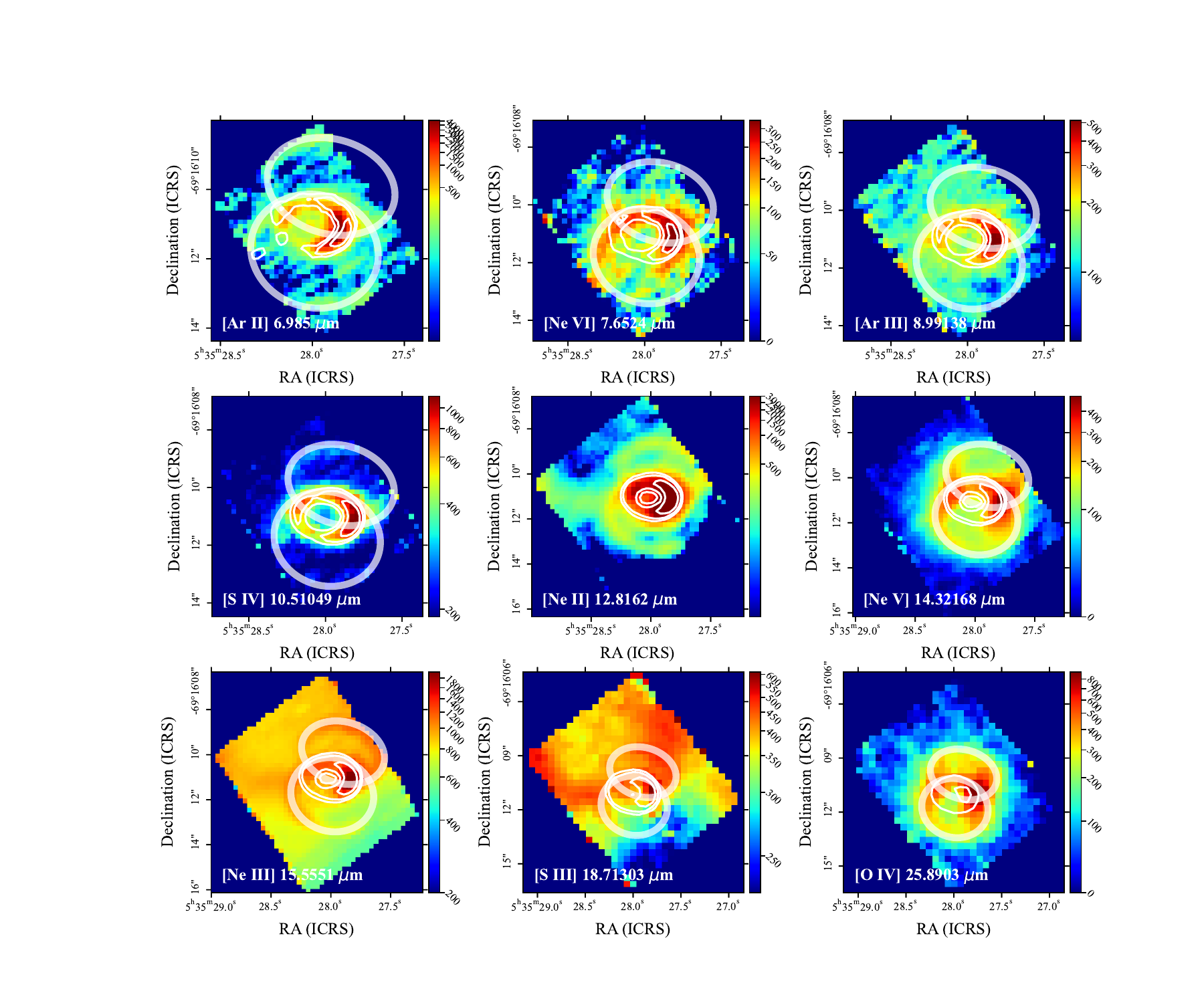}
\caption{Continuum subtracted emission line images for those species with emission from outside the ER. The grey circles overlaid on the images are the locations of the outer rings determined from the [Ne~{\sc ii}]~12.8135~\micron\ morphology, with the light-grey contours showing the ER continuum morphology determined from an `off' position adjacent to the line. The colour bars are in units of MJy/sr.}
\label{Fig:outside_temp}
\end{figure*}

\subsection{Line emission from the Equatorial Ring and surrounds}
\label{sec:lines}

\subsubsection{Emission line spectral properties}
\label{sect:emission_line_props}
To investigate the emission line properties of the ER and its adjacent regions, we analyzed the spectrum corresponding to the whole ring, shown in Figure~\ref{Fig:spec}, as well as the spectra extracted for the four cardinal point sub-regions of the ER that are shown in the lower part of Fig.~\ref{Fig:spec_extraction_Regions}. Emission line radial velocities, full-width half-maximum (FWHM) values and integrated line fluxes were measured using the Gaussian emission line fitting (ELF) routines written by P. J. Storey within the {\sc dipso}
spectral analysis package
\citep{Howarth2014}. Results are presented in Table~\ref{tab:whole_ring} for the whole ER spectrum and in Table~\ref{tab:cardinal_points} for the four cardinal point spectra. The solid angle corresponding to each spectral extraction is listed in the table headers.

Sixteen emission line detections are listed in each table. H~{\sc i} 8-6 7.502~$\mu$m was also detected. In addition, a very broad 26-$\mu$m feature with multiple peaks (FWZI$\sim$8000~km~s$^{-1}$) is present in each of the spectra, corresponding to [Fe~{\sc ii}] 25.98839~$\mu$m, superposed on what is a narrow [O~{\sc iv}] 25.8903~$\mu$m line (see Fig.~\ref{Fig:spec}). A weaker blueshifted component to this feature is discussed in Section~\ref{sec:ejecta} and shown in Figure~\ref{Fig:fe_ii26_534_neii_comp}. For the whole ER spectrum  the integrated [Fe~{\sc ii}] 25.99 $\mu$m line flux is measured to be $(2.3\pm0.1)\times10^{-21}$~W~cm$^{-2}$, similar to the flux levels measured for this line from a time sequence of {\em Spitzer} spectra obtained up to day 7954 (year 21.8) \citep{Arendt2016}. A broad redshifted emission line corresponding to [Fe~{\sc ii}] 5.340169~$\mu$m is also present in the whole ER spectrum, with a flux of $(6.0\pm0.25)\times10^{-22}$~W~cm$^{-2}$, a heliocentric line centre radial velocity of +2450$\pm$30~km~s$^{-1}$ and an  FWHM of 1600$\pm$70~km~s$^{-1}$.
The only cardinal point spectrum in which the 5.34 $\mu$m line is detected is the Ring-South spectrum (flux=$(6.2\pm0.13)\times10^{-23}$~W~cm$^{-2}$, velocity$=+2990\pm11$~km~s$^{-1}$, FWHM$=1055\pm27$~km~s$^{-1}$). These lines are due to the ejecta-ring interaction and are discussed further in Section~\ref{sec:ejecta}.

A comparison of the MRS line fluxes for the whole ER listed in Table~\ref{tab:whole_ring}
with those listed by \citet{Arendt2016} from their {\em Spitzer} 10--40 $\mu$m spectral sequence shows the latter fluxes to be generally an order of magnitude larger. This might be due to the {\em Spitzer} line flux measurements having been made with larger apertures. A comparison of the relative line intensity patterns seen in the {\em JWST}-MRS and {\em Spitzer}-IRS datasets indicates that, relative to [S~{\sc iv}] 10.51 $\mu$m, the low degree of ionization [Fe~{\sc ii}] and [Ne~{\sc ii}] lines have brightened by factors of 5-12 in the MRS spectra. \citet{Arendt2016} suspected the presence of [Ni~{\sc ii}] 6.64~$\mu$m, H~{\sc i} 7.46~$\mu$m and [Ne~{\sc vi}] 7.65~$\mu$m in their R=90 5--10 $\mu$m {\em Spitzer} spectra - these lines are confirmed by the MRS spectra. The MRS detection of [Ar~{\sc iii}] 8.99~$\mu$m appears to be new. The other new line detections are [Fe~{\sc ii}] 5.340169~$\mu$m, [Ni~{\sc ii}] 10.6822~$\mu$m and [Na~{\sc iii}] 7.31775~$\mu$m, both seen in the whole-Ring and in the Ring-West spectra.

The 10--40 $\mu$m {\em Spitzer}-IRS spectra discussed by \citet{Arendt2016} had a resolving power of R=600. The $\sim$5 times greater resolving power of the 5--27 $\mu$m {\em JWST}-MRS spectra allows higher resolution FWHM and radial velocity measurements to be made. Inspection of the FWHM values for the lines in the whole ER spectrum listed in Table~\ref{tab:whole_ring} reveals two groupings: (a) eight lines with FWHM values of between 280 and 380~km~s$^{-1}$, namely all the lines that arise from singly ionized species (Ni$^+$, Ar$^+$, H$^+$, Ne$^+$ and Fe$^+$), together with lines from doubly ionized Na$^{2+}$ and Ar$^{2+}$; and (b) seven lines arising mainly from more highly ionized species (Ne$^{5+}$, Ne$^{4+}$, Ne$^{2+}$, S$^{3+}$, S$^{2+}$ and O$^{3+}$), with FWHM values between 99 and 171~km~s$^{-1}$. The [S~{\sc iv}] 10.51 $\mu$m, [Ne~{\sc vi}] 7.65 $\mu$m and [Ne~{\sc v}] 14.32 $\mu$m lines are particularly
narrow (99--107~km~s$^{-1}$ FWHM) - not much larger than the instrumental resolution measured at those wavelengths by \cite{Jones2023}, while the larger line widths (123--145~km~s$^{-1}$) measured for the [S~{\sc iii}] 18.71 $\mu$m, [Ne~{\sc v}] 24.32 $\mu$m and [O~{\sc iv}] 24.89 $\mu$m lines are consistent with the instrumental resolutions measured by \citet{Jones2023} for this lower resolution part of the MRS range.

The general expansion of the ER may contribute to the broadening of some of the lines seen in the whole-ER spectrum, an effect which should be reduced in the individual cardinal point spectra. Comparison of the FWHM values measured for the whole ER (Table~\ref{tab:whole_ring}) with those measured for the cardinal point spectra (Table~\ref{tab:cardinal_points})
shows that for the singly ionized species their FWHM values are appreciably lower in the cardinal point spectra. However, this is not the case for the much narrower lines that arise from the more highly ionized species, for which the whole-ER and cardinal point FWHM values are similar. This behaviour indicates that while the emission from the singly and doubly ionized species originates from the ER itself, that from the more highly ionized species may have a partly different origin. This is also shown by the line maps presented in Section~\ref{sect:emission_line_maps}. Support for this interpretation comes from a consideration of the radial velocity differences seen in Table~\ref{tab:cardinal_points} across the four cardinal point spectra. For lines from singly ionized species, the mean difference between the radial velocities measured for the Ring-South and Ring-North spectra was +286.7$\pm$45.5~km~s$^{-1}$, while the East-West mean difference was +46.9$\pm$29.5~km~s$^{-1}$. In contrast, for the more highly ionized species, the corresponding radial velocity differences were much smaller, +45.7$\pm$49.1~km~s$^{-1}$ (South-North) and $-7.1\pm$10.0~km~s$^{-1}$ (East-West). We take this as confirmation that the lines from singly and doubly ionized species originate principally from the expanding ER and that the much narrower lines from more ionized species must originate from a different emission component.

\citet{Groningsson2008} presented 6~km~s$^{-1}$ resolution optical spectra of northern and southern parts of the ER, obtained in 2002 at days 5702--5705 (year 15.6) past explosion, at a time when the northern quadrant was the brightest part of the ER. They saw a mixture of narrow lines (FWHM=10-29~km~s$^{-1}$), from H~{\sc i} up to [Ne~{\sc v}], and broad lines (line widths larger than 200~km~s$^{-1}$), from H~{\sc i} up to [Fe~{\sc xiv}]. They attributed the narrow lines to unshocked pre-ionized gas in the ER and the broad lines to recombining post-shock gas. By the time of our MIRI-MRS observations of the ER on day 12927 (year 35.4), the only broad emission lines seen were from the singly ionized species, consistent with the ongoing recombination of the post-shock gas and with \citet{Groningsson2008}'s prediction that ``As more and more gas cools we expect the width of especially the low ionization lines to increase.''

From optical emission line diagnostics and thermally broadened line widths, \citet{Groningsson2008} estimated
electron densities $n_e \sim 1500$--5000~cm$^{-3}$ and electron temperatures $T_e \sim 6500$--24,000~K for their narrow-line component. The narrow-line [Ne~{\sc v}] 14.32/24.32 $\mu$m line flux ratio is the only electron density diagnostic available from our MRS spectra of the ER. For the Ring-West spectrum, the brightest of the four cardinal point spectra (see Table~\ref{tab:cardinal_points}), this ratio is measured to be 1.151$\pm$0.115, while for the whole-ER spectrum (see Table~\ref{tab:whole_ring}) the ratio is found to be 1.021$\pm$0.059. Using the atomic data of \citet{Galavis1997} and \citet{Griffin2000}, these ratios correspond to electron densities $n_e$ of between 700 and 4300~cm$^{-3}$ for electron temperatures in the range $T_e = 5000-25000$~K, similar to the parameters derived by \citet{Groningsson2008} for their narrow line component.

The detection of two [Ni~{\sc ii}] lines in the Ring-West spectrum allows a constraint to be placed on the electron temperature of this part of the ER. Using the atomic data of \citet{Nussbaumer1982}, \citet{Quinet1996} and \citet{Bautista2004}, the observed [Ni~{\sc ii}] 6.64/10.68~$\mu$m flux ratio of 9.71$\pm$0.35 requires an electron temperature T$_e \leq 6500$~K. For temperatures higher than this the 10.68 $\mu$m line (which originates from a higher level than the upper level of the 6.64 $\mu$m transition) would become too strong. This temperature limit is consistent with strong post-shock cooling having taken place in the dense ER clumps. 

\subsubsection{Line emission maps}
\label{sect:emission_line_maps}

Here, we discuss the spatially resolved line emission structure of SN~1987A over the MRS IFS for the different narrow- and broad-line-emitting species.
For each selected emission line, we formed continuum-subtracted images by subtracting from each line-centred image a mean image  from adjacent line-free regions on either side of the line, each off-image having the same spectral bandwidth as the line-centred image. 

Fig.~\ref{Fig:outside_temp} shows continuum-subtracted images for nine emission lines, using display ranges designed to bring out fainter levels of emission. 
All of the images show strong ER emission, peaking at the western quadrant. The outer rings are prominent in the [Ne~{\sc ii}], [Ne~{\sc iii}],  
[Ne~{\sc v}], [Ne~{\sc vi}] and [O~{\sc iv}] images, with the observed distributions of the latter three ions predominantly following the morphology of the outer rings.
All of the line images shown in Fig.~\ref{Fig:outside_temp} display some level of extended emission beyond the extent of the inner and outer rings. For [Ne~{\sc v}] and [O~{\sc iv}] this lower surface brightness emission appears to closely envelop the outer rings but for [Ne~{\sc ii}], [Ne~{\sc iii}] and, in particular, [S~{\sc iii}] 18.71~$\mu$m, part of a large ring-like structure fills the north-eastern half of the IFS FoV, passing through the northern part of the ER. In order to check for radial velocity variations across this structure that might support a very large expanding ring interpretation, we extracted spectra of the [S~{\sc iii}] 18.71 $\mu$m line at three positions in the ring-like structure: one from near its north-eastern edge; one near its
north-western edge; and one just to the north of the ER. However, the measured [S~{\sc iii}] radial velocities showed a difference of only 4.9$\pm$0.7~km~s$^{-1}$ between the NE and NW positions, with the velocity of the position north of the ER measured to be in between those of the other two. We interpret these results as not providing strong support for a very large expanding ring interpretation.

\begin{figure*}
\centering
\includegraphics[width=0.98\hsize,trim={1.5cm 1.5cm 1.5cm 1.5cm},clip]{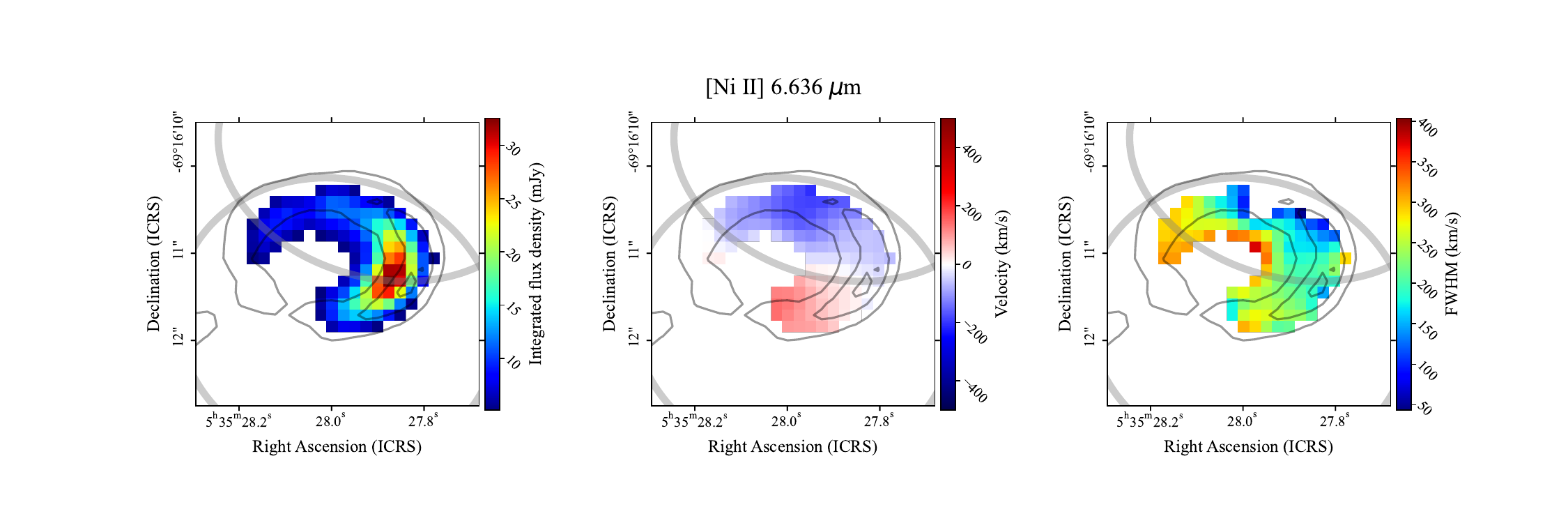}
\includegraphics[width=0.98\hsize,trim={1.5cm 1.5cm 1.5cm 1.5cm},clip]{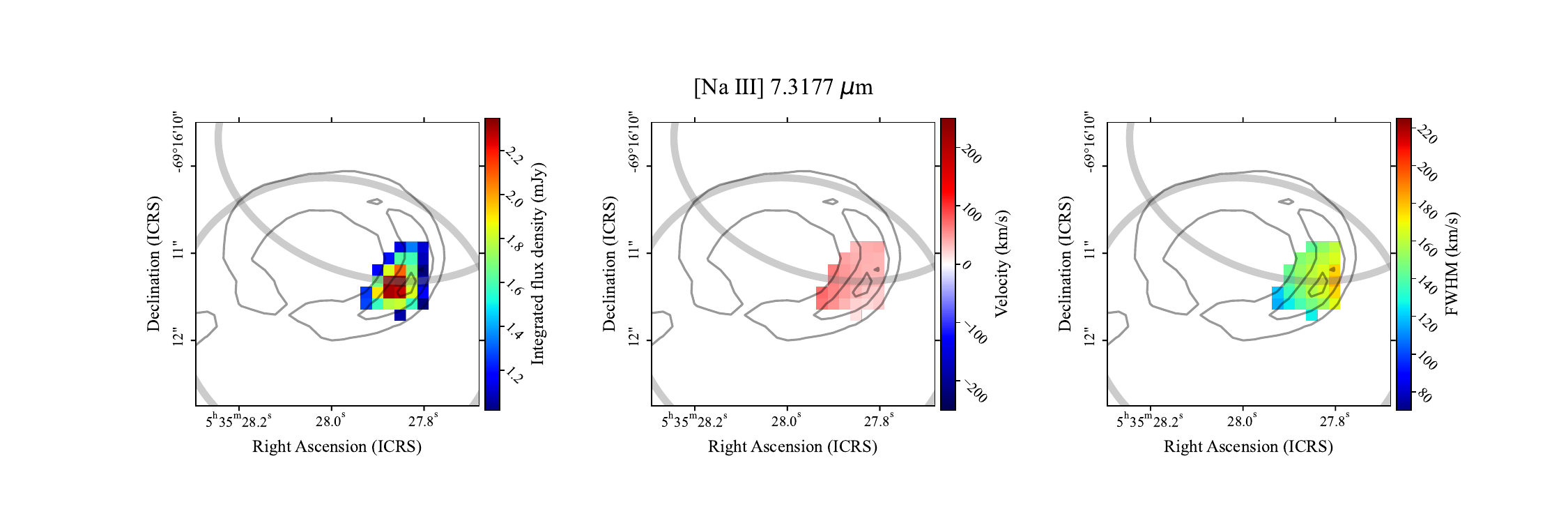}
\includegraphics[width=0.98\hsize,trim={1.5cm 1.5cm 1.5cm 1.5cm},clip]{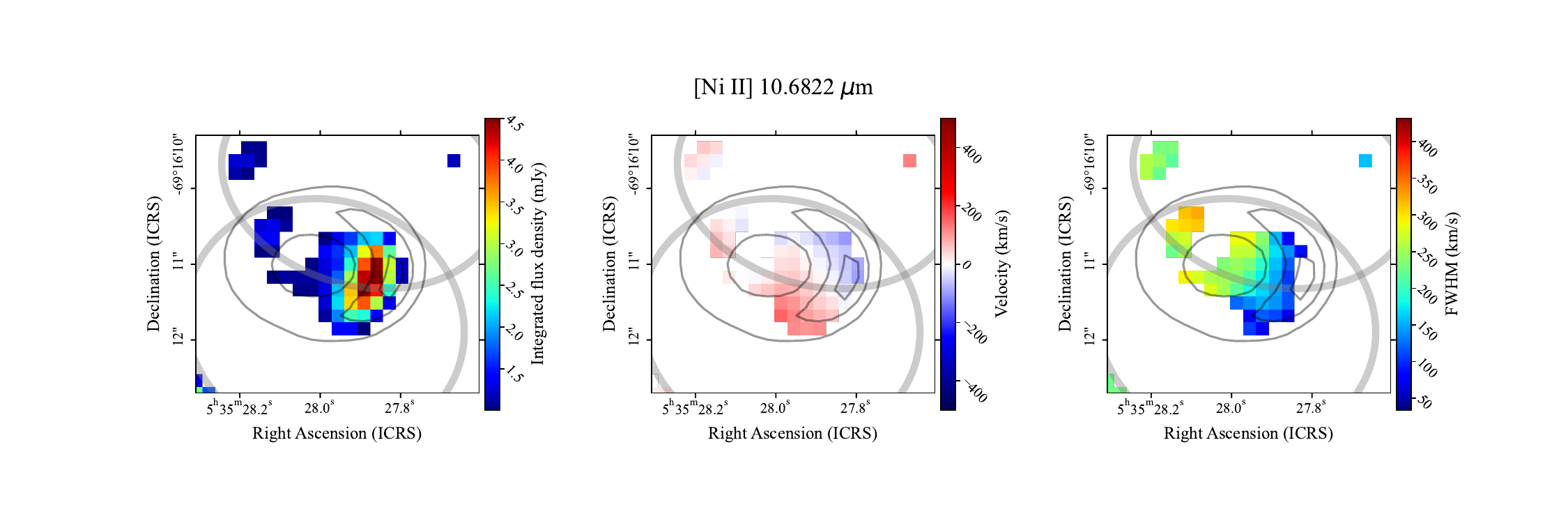}
\includegraphics[width=0.98\hsize,trim={1.5cm 1.5cm 1.5cm 1.5cm},clip]{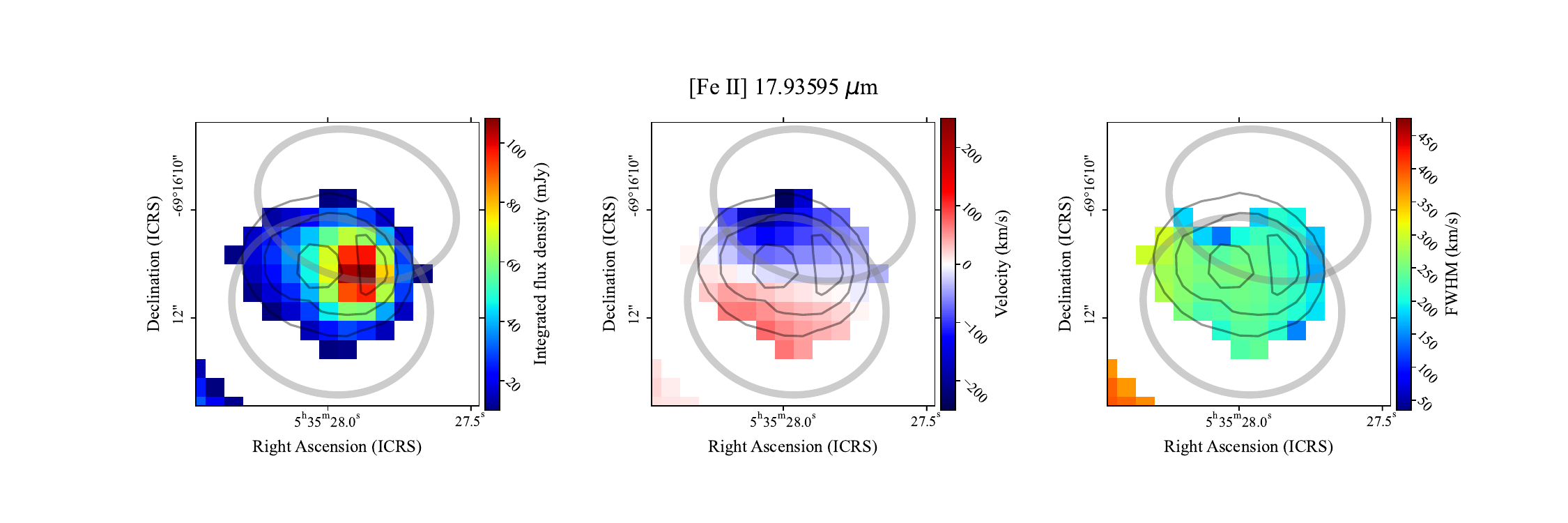}
\caption{ Moment maps for a selection of species prominent in the ring ordered row-by-row. The left panes show the integrated line intensity. The middle panes show the radial velocity of the line emission, corrected for the systemic barycentric velocity of SN~1987A
of +287~km~s$^{-1}$ \citep{Groningsson2008}. The right panes show the line FWHMs. The grey circles overlaid on the images are the locations of the outer rings derived determined from the [Ne~II]~12.8135~\micron\ morphology, with the grey contours showing the ER continuum morphology determined from an `off' position adjacent to the line.}
\label{Fig:fe_ni_ring_mm}
\end{figure*}

\begin{figure*}
\centering
\includegraphics[width=0.95\hsize,trim={1.5cm 1.5cm 1.5cm 1.5cm},clip]{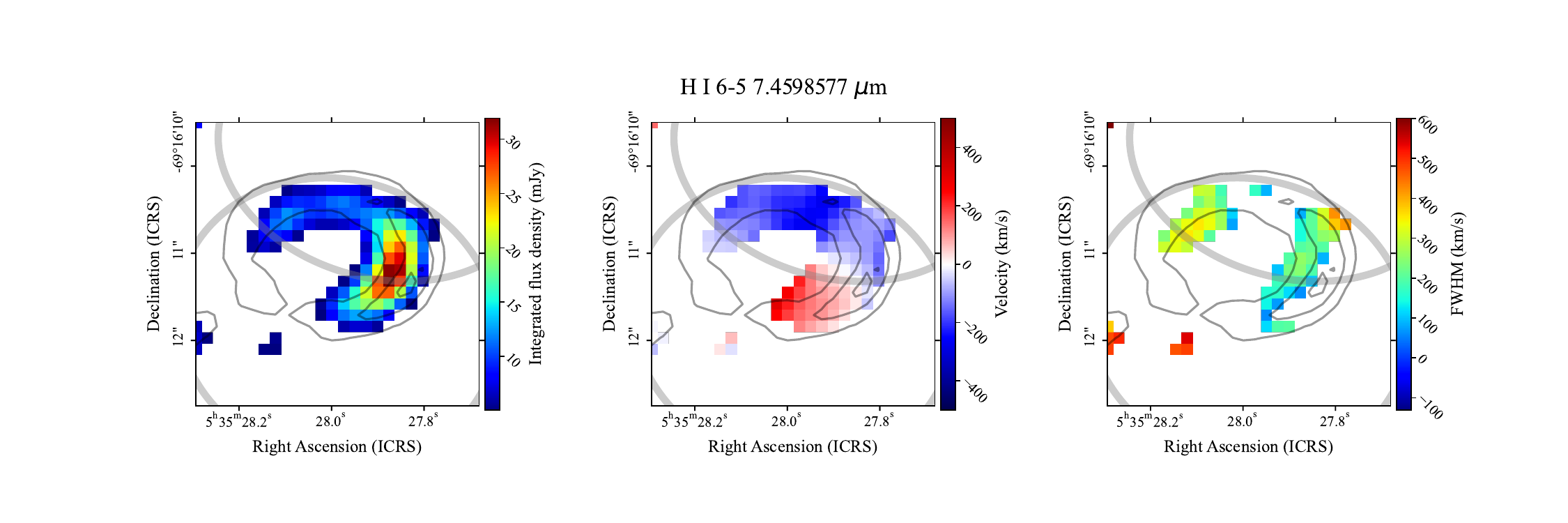}
\includegraphics[width=0.95\hsize,trim={1.5cm 1.5cm 1.5cm 1.5cm},clip]{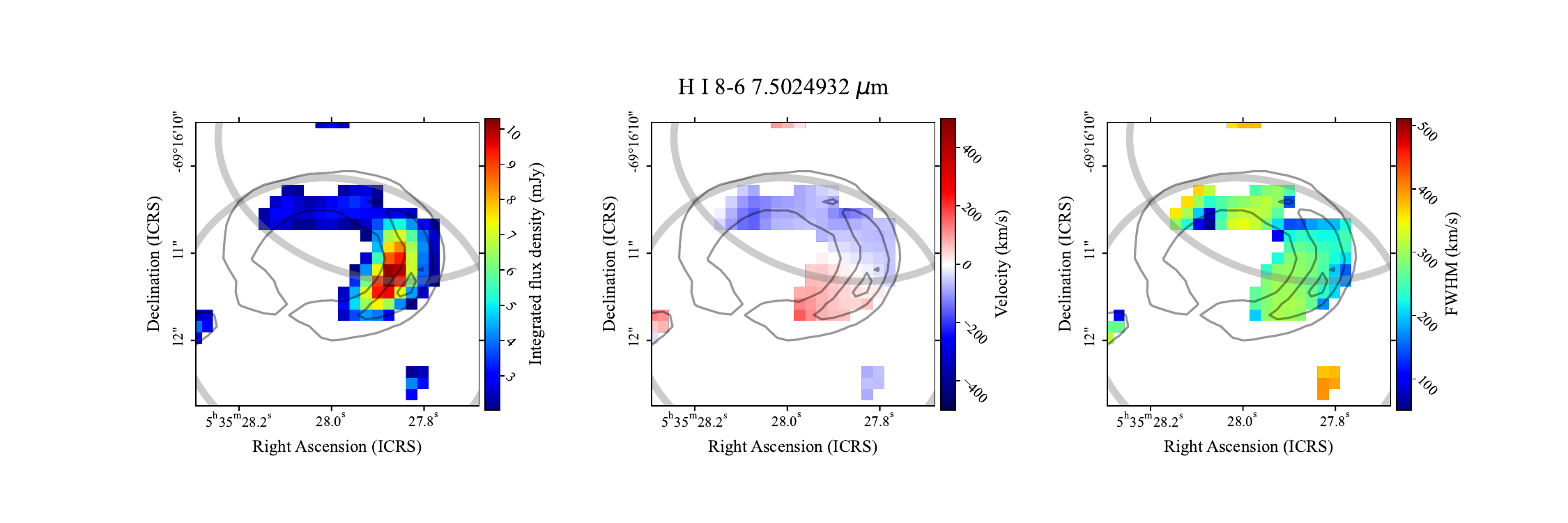}
\caption{Same as Fig.\ref{Fig:fe_ni_ring_mm} but for a selection of H~I lines.}
\label{Fig:hi_ring_mm}
\end{figure*}

\begin{figure*}
\centering
\includegraphics[width=0.95\hsize,trim={1.5cm 1.5cm 1.5cm 1.5cm},clip]{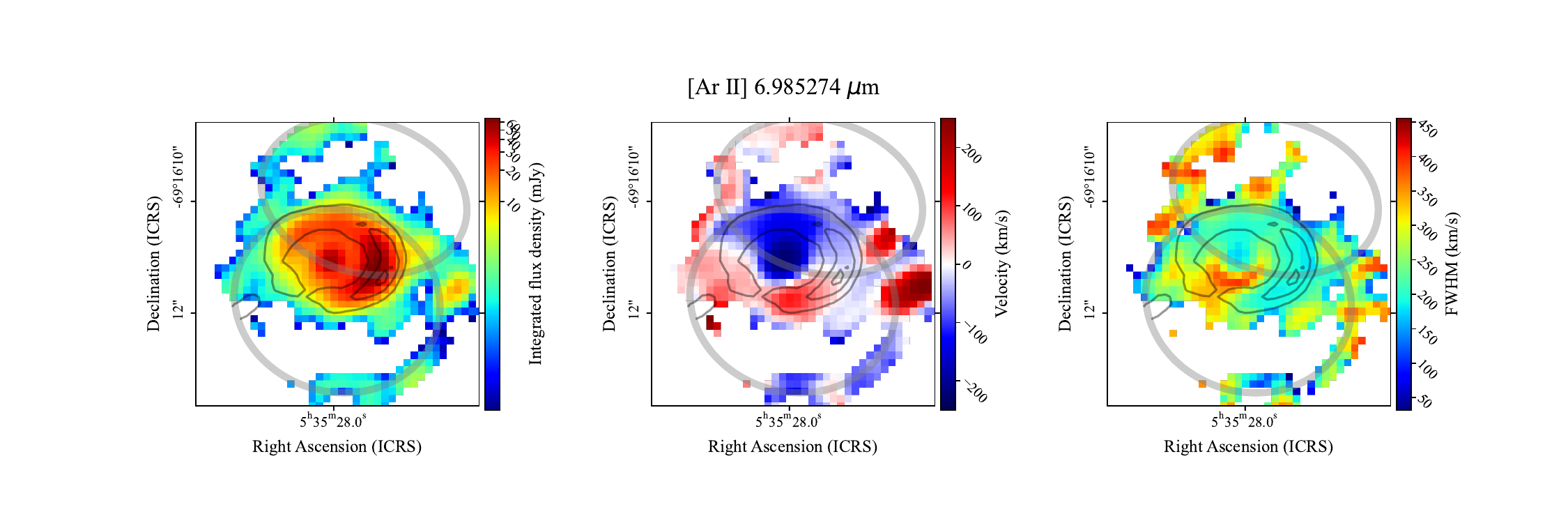}
\includegraphics[width=0.95\hsize,trim={1.5cm 1.5cm 1.5cm 1.5cm},clip]{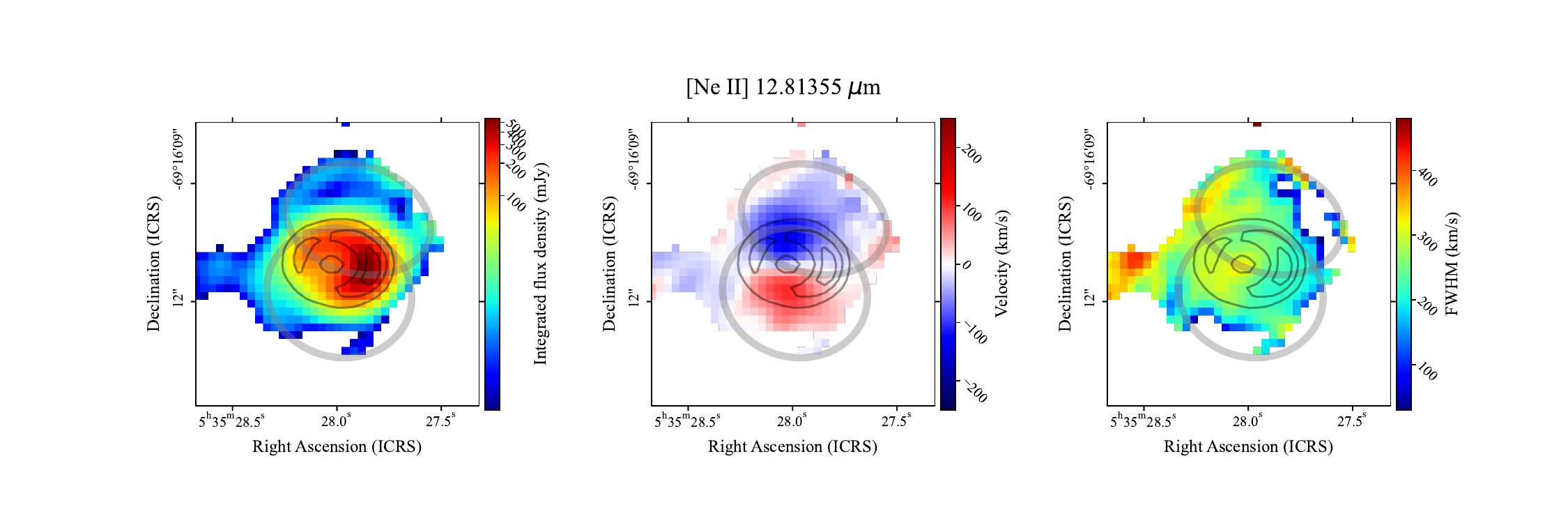}
\includegraphics[width=0.95\hsize,trim={1.5cm 1.5cm 1.5cm 1.5cm},clip]{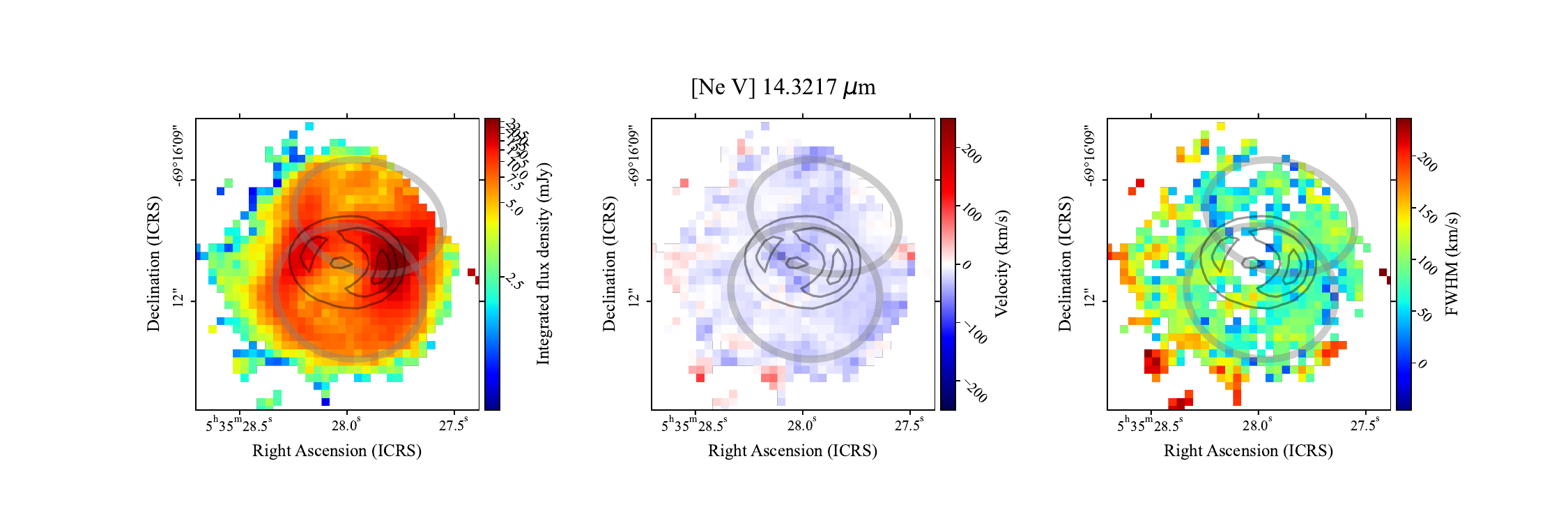}
\caption{Same as Fig.\ref{Fig:fe_ni_ring_mm} but for a selection of Ar and Ne lines that reveal material in the outer rings and outside the ER.}.
\label{Fig:ar_ne_ring_mm}
\end{figure*}

We have identified spectral features of both small and large spatial scales by manually searching through the spectral axis MRS sub-band datacube using small and large apertures at varying locations around the system. We then identified the emitting species associated with each feature before slicing a sub-cube around the feature using the Spectral-Cube\footnote{\url{https://spectral-cube.readthedocs.io/en/latest/}} Astropy affiliated package \citep{Ginsburg2014}. We carefully fitted and subtracted the continuum from each of the sub-cube spaxels using regions away from the spectral feature and accounting for the variation in sampling and residual fringing. With the resulting sub-cubes, we created sets of moment maps using Spectral-Cube. From the 0th, 1st, and 2nd moments, we produced integrated line intensity maps, velocity maps which account for the systemic barycentric velocity of SN~1987A of 287~km~s$^{-1}$ \citep{Groningsson2008}, and FWHM maps. We applied a cut across all maps based on an integrated intensity threshold to mask regions with little or no line emission. The moment maps for a selection of lines are shown in Figs.~\ref{Fig:fe_ni_ring_mm}, ~\ref{Fig:hi_ring_mm} and ~\ref{Fig:ar_ne_ring_mm}. Lines from singly ionized species, the majority of those plotted in the three figures, all show a strong radial velocity difference between the blue-shifted northern part of the ER and the red-shifted southern part, consistent with the cardinal point radial velocity measurements for singly ionized species listed in Table~\ref{tab:cardinal_points},
and with previous ground-based results \citep[e.g.][]{Kjaer2007}. 

The radial velocity map shown in Fig.~\ref{Fig:ar_ne_ring_mm} for [Ne~{\sc v}] 14.32~$\mu$m, a representative of the more highly ionized species, does not show any radial velocity changes across the ER, consistent with its velocity measurements in Table~\ref{tab:cardinal_points}, which show differences
of only 2.0$\pm$1.8 and 13.6$\pm$2.5~km~s$^{-1}$ between its East-West and South-North radial velocities, respectively.
But Table~\ref{tab:cardinal_points} also shows that the fluxes measured for [Ne~{\sc v}] and other high ionization species peak in the Ring-West aperture, indicating a relationship between the narrow-line component and the ER material.
However, the narrow-line component also extends to the outer rings and beyond, up to the edges of the MRS FoV - see Figs.~\ref{Fig:outside_temp} and \ref{Fig:ar_ne_ring_mm} for e.g. [Ne~{\sc v}], with the mean surface brightness of the [Ne~{\sc v}] 14.32-$\mu$m line in the background aperture (Fig.~\ref{Fig:spec_extraction_Regions}) corresponding to 4~per cent of its surface brightness in the Ring-West aperture.

\subsection{Dust in the Equatorial Ring}
\label{sec:ER_dust}

\subsubsection{Dust Composition}

The MRS spectrum of the ER, shown in Figure~\ref{Fig:spec}, is dominated by continuum emission from amorphous silicate dust, characterised by the broad $\sim$10~$\mu$m and $\sim$18~$\mu$m spectral features. 
No evidence for polycyclic aromatic hydrocarbons (PAH) emission is present in the spectrum; the improved resolution of the MRS has not revealed any new solid-state dust species in the mid-IR spectra compared to {\em Spitzer}/IRS observations. 

\begin{figure*}
\centering
\includegraphics[width=0.6\hsize]{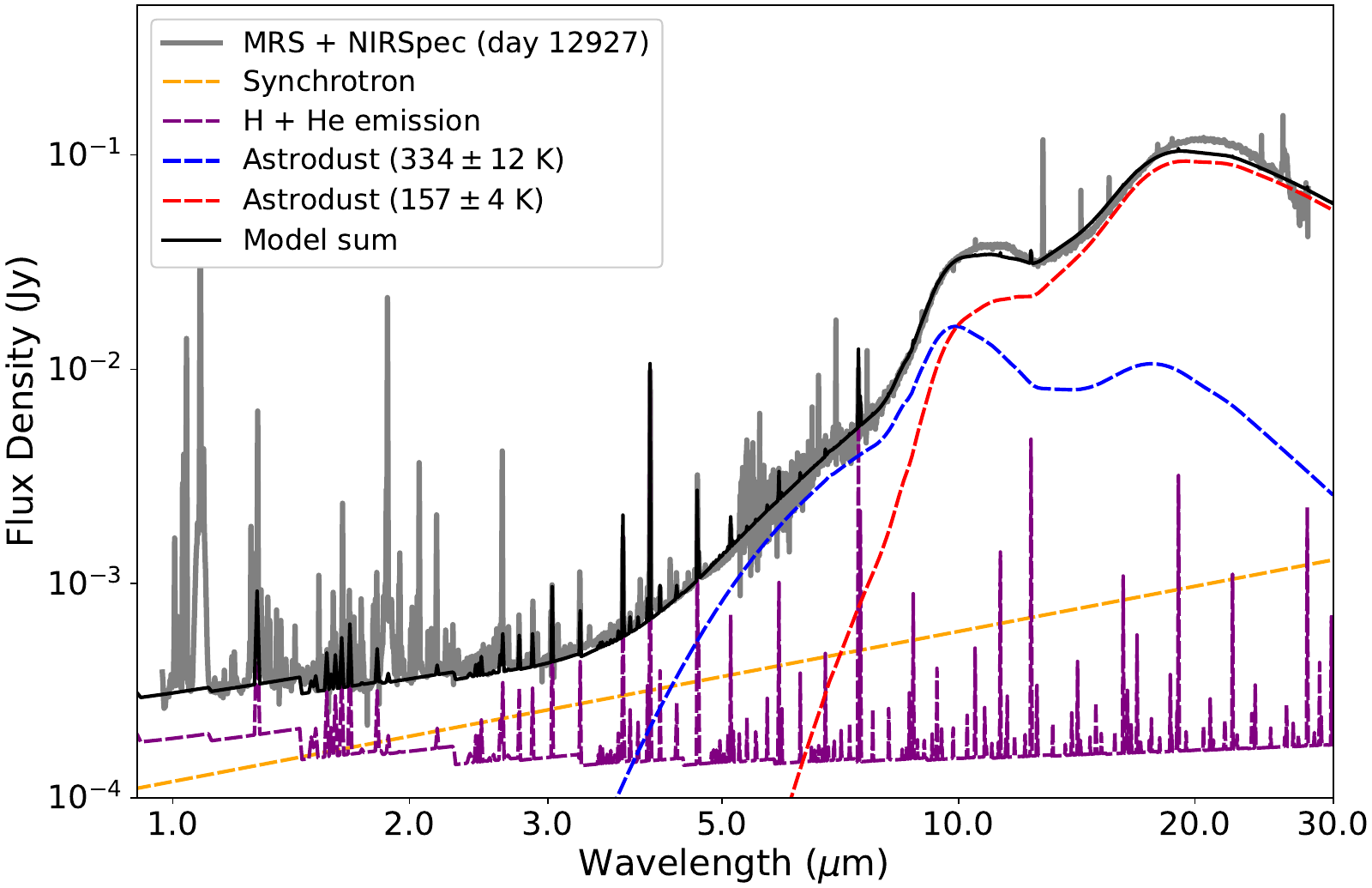}
\includegraphics[width=0.6\hsize]{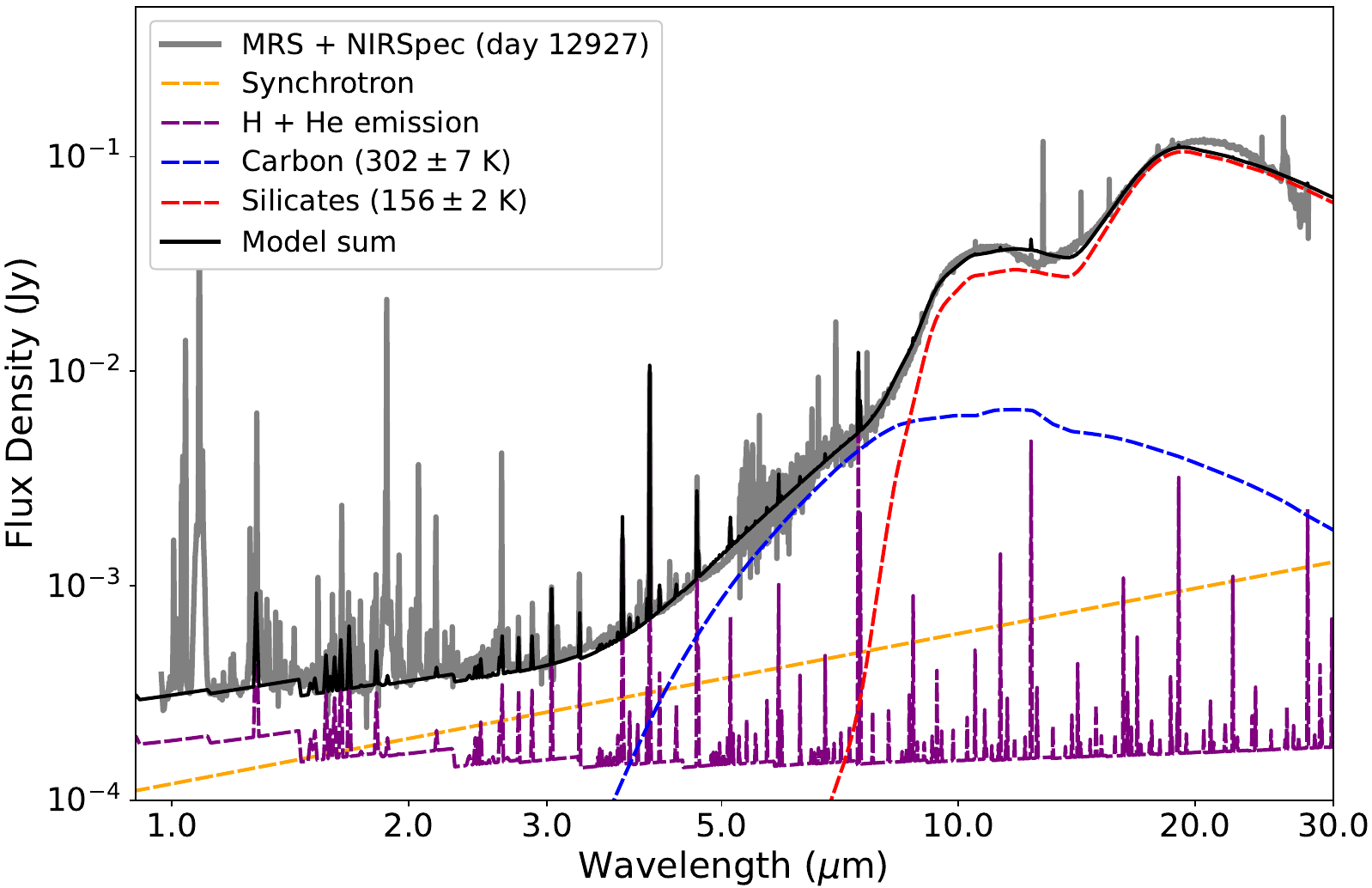}
\includegraphics[width=0.6\hsize]{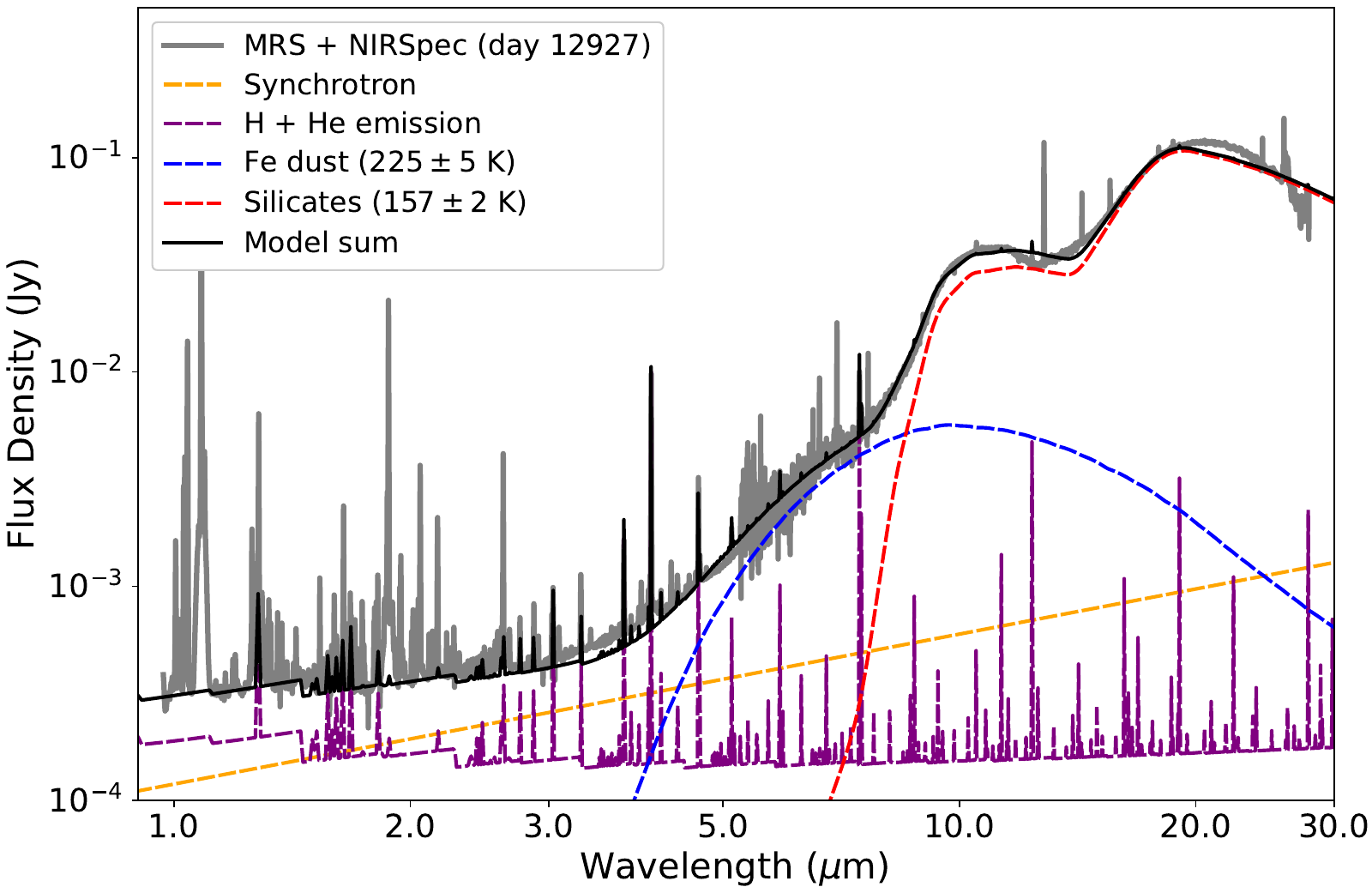}
\caption{Fits to the MRS and NIRspec data using the two-temperature dust compositions of \textit{astrodust} \citep{Hensley2023} in the top panel, carbon and silicates \citep{Draine1984} in the middle panel, and silicates \citep{Draine1984} and iron dust in the bottom panel. The synchrotron emission (orange line) has been extrapolated from ALMA observations based on the models described in \citet{Cendes2018} and \citet{Cigan2019} (see text for details). The emission from the H and He continuum and lines (purple line) is described in \citet{Larsson2023}.  The best-fit parameters for the dust components are summarized in Table~\ref{Table:dusttab}.} 
\label{Fig:astrodust_fit}
\end{figure*}

In order to determine which dust properties best reproduce the shape of the dust continuum observed with \textit{JWST}, we fitted the 5.3--28~$\micron$ MRS and the 0.9--5.3~$\micron$ NIRSpec data \citep{Larsson2023} with two-temperature dust models of varying grain compositions. We accounted for the synchrotron emission for day 12,927 using the model constructed from ALMA observations of the ER that is described by Equation~1 of \citet{Cendes2018}. We used the parameters from the torus model listed in Table~6 of \citet{Cendes2018} and the updated value of the spectral index $\alpha$ of 0.7 from \citet{Cigan2019}. The parameters for the synchrotron component were fixed and did not vary in our fit. We also included bound-free and free-free emission from \ion{H}{1}, \ion{He}{1}, and \ion{He}{2} with a temperature of $10^4$~K that was used by \citet{Larsson2023} to fit the NIRSpec data. The models are shown in Figure~\ref{Fig:astrodust_fit} and the best-fit temperatures and corresponding dust masses are listed in Table~\ref{Table:dusttab}. 

We find that the 0.9--28~$\micron$ spectral range is best-reproduced by a single grain composition of \textit{astrodust} \citep{Draine2021, Hensley2023} emitting at 157$\pm$4 and 334$\pm$12~K. This model is shown in the top panel of Figure~\ref{Fig:astrodust_fit}. As described in \citet{Hensley2023}, \textit{astrodust} reproduces the observational properties of dust in the diffuse ISM and is made up of composite grains that consist of different compositions (primarily silicates and oxides) on small scales but are effectively one uniform composition on larger scales of $>$0.05~$\micron$. We assumed that the grains are oblate spheroids with an axial ratio of 1.4:1, a porosity of 0.2, and a density of 2.74~g\:cm$^{-3}$ \citep[see][]{Hensley2023}. Since the mass absorption coefficients are size-independent in the mid-IR wavelength range, the choice of grain size did not affect our best-fit parameters.  
 
We also attempted to fit the spectra using a silicate dust grain composition \citep{Draine1984}, but as was found in earlier studies \citep[e.g.][]{Dwek2010}, we required an additional featureless dust component to account for the shorter wavelength emission. For this second component, we used carbon dust from \citet{Draine1984} (middle panel of Figure~\ref{Fig:astrodust_fit}) and pure iron dust from \citet{Ordal1985} (bottom panel). Both combinations fit the spectra fairly well, even though the fits are slightly poorer than the \textit{astrodust} in the 3--5~$\micron$ range. The corresponding dust masses and temperatures are listed in Table~\ref{Table:dusttab}. The total dust mass required from the pure Fe dust component is unreasonably high, so we disfavour this composition as a dominant component producing the shorter-wavelength dust continuum.

The appropriateness of the {\em astrodust} ISM dust model of \cite{Draine2021} for the ER dust around SN~1987A is a valid issue, given that its immediate progenitor was a B supergiant. While `normal' B supergiants do not form dust in their outflows \citep{Barlow1977}, some luminous blue variables with B supergiant spectra have circumstellar dust shells that exhibit mid-IR silicate emission bands \citep[e.g.][]{Voors2000}, but existing observations lack sufficient signal-to-noise for comparisons to be made to the {\em astrodust} model. Another possibility is that the ER dust might have been produced during an earlier M supergiant phase of the progenitor star. The mid-IR spectral survey of M supergiants by \cite{Speck2000} found a wide variety of 10-$\mu$m silicate feature shapes, including a `broad feature' group whose 10-$\mu$m profiles exhibit similarities to that of SN~1987A's ER. The need for additional opacity shortwards of 7-$\mu$m in cool star dust envelope models, over and above that provided by pure silicates, has long been recognised, leading to the introduction of `dirty silicate' models for such sources \citep[e.g.][]{Jones1976}.

\begin{figure}
\centering
\includegraphics[width=\hsize]{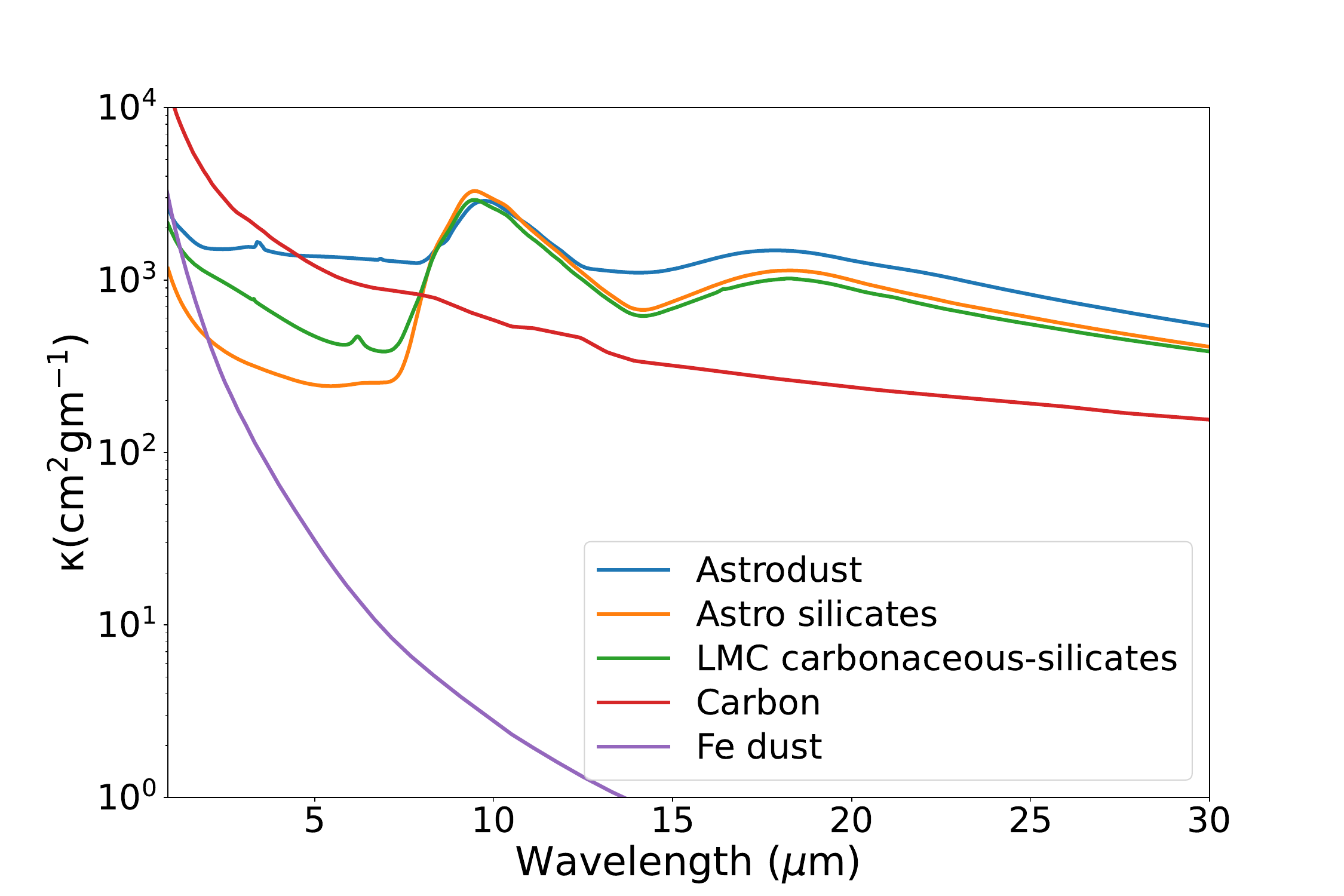}
\caption{Dust mass absorption coefficients for some of the common grain species. The \textit{astrodust} from \citet{Hensley2023} is shown in blue, silicates from \citet{Draine1984} in orange, LMC carbonaceous-silicates from \citet{Weingartner2001} in green, carbon in red \citep{Draine1984}, and iron grains in purple.} 
\label{Fig:kappas}
\end{figure}

Out of the grain compositions that we used, only the \textit{astrodust} composition could reproduce the observed spectrum without the need for an additional featureless grain composition. The reason for this can be seen in the comparison of the mass absorption coefficient ($\kappa$) values for each of the grain species, shown in Figure~\ref{Fig:kappas}. It can be seen that the \textit{astrodust} opacity is significantly larger in the 2.5--7.0~$\micron$ range, and the peak-to-continuum ratio for the broad 9.8 and 20~$\micron$ features lower, compared to silicates from \citet{Draine1984} and the LMC carbonaceous-silicate grains from \citet{Weingartner2001}.

\begin{deluxetable}{lcc}
\tablecolumns{3} \tablewidth{0pc} \tablecaption{\label{Table:dusttab}Dust Model Fit Parameters}
\tablehead{
\colhead{Dust} & \colhead{Temperature} & \colhead{Mass} \\
\colhead{Composition}  & \colhead{(K)} & \colhead{($\rm 10^{-5} \:M_{\odot}$)}
}
\startdata
Astrodust & 334 $\pm$ 12 &  $0.012\pm0.003$ \\
Astrodust & 157 $\pm$ 4 &  $1.7\pm0.2$ \\
\hline
Carbon & 302 $\pm$ 7 &  $0.037\pm0.007$ \\
Silicates & 156 $\pm$ 2 &  $2.5\pm0.2$ \\
\hline
Iron & 225 $\pm$ 5 &  $35.1\pm8.4$ \\
Silicates & 157 $\pm$ 2 &  $2.4\pm0.2$ \\
\enddata
\tablecomments{Dust temperatures and masses for the models shown in Figure~\ref{Fig:astrodust_fit}. The listed uncertainties are the 3-$\sigma$ uncertainties from the fit.}
\end{deluxetable}



\begin{deluxetable}{ccccc}
\tablecolumns{5} \tablewidth{0pc} \tablecaption{\label{Table:tempdust}Temporal Evolution in Dust Properties}
\tablehead{
\colhead{Day} & \colhead{T$_1$} & \colhead{M$_1$} & \colhead{T$_2$} & \colhead{M$_2$} \\
\colhead{}  & \colhead{(K)} & \colhead{($10^{-8}\: \rm M_{\odot}$)} & \colhead{(K)} & \colhead{($10^{-5}\: \rm M_{\odot}$)}}

\startdata
6805 & $383^{+29}_{-23}$ & $7.8^{+3.7}_{-2.8}$ & $162^{+7}_{-7}$ & $0.85^{+0.15}_{-0.12}$\\
7138 & $394^{+21}_{-18}$ & $8.2^{+2.5}_{-2.1}$ & $162^{+5}_{-5}$ & $1.1^{+0.1}_{-0.1}$ \\
7296 & $393^{+21}_{-18}$ & $8.2^{+2.6}_{-2.1}$ & $162^{+5}_{-5}$ & $1.1^{+0.1}_{-0.1}$ \\
7555 & $374^{+22}_{-18}$ & $11.2^{+4.6}_{-3.6}$ & $164^{+5}_{-5}$ & $1.2^{+0.2}_{-0.1}$ \\
7799 & $370^{+23}_{-19}$ & $13.7^{+5.6}_{-4.3}$ & $162^{+6}_{-6}$ & $1.4^{+0.2}_{-0.2}$ \\
7955 & $382^{+44}_{-11}$ & $11.0^{+8.4}_{-5.2}$ & $164^{+8}_{-9}$ & $1.4^{+0.3}_{-0.2}$ \\
12927 & $375^{+4}_{-4}$ & $6.26^{+0.45}_{-0.43}$ & $161^{+1}_{-1}$ & $1.50^{+0.03}_{-0.03}$ \\
\enddata
\tablecomments{Dust temperatures and masses for two-component fits to the 5-30~$\micron$ spectra from 
\textit{Spitzer} (days 6805--7955) and {\em JWST} MRS (day 12927) using the \textit{astrodust} composition from \citet{Hensley2023}. The listed uncertainties are the 3-$\sigma$ uncertainties from the fit.}
\end{deluxetable}


\begin{figure*}
\centering
\includegraphics[width=0.48\hsize]{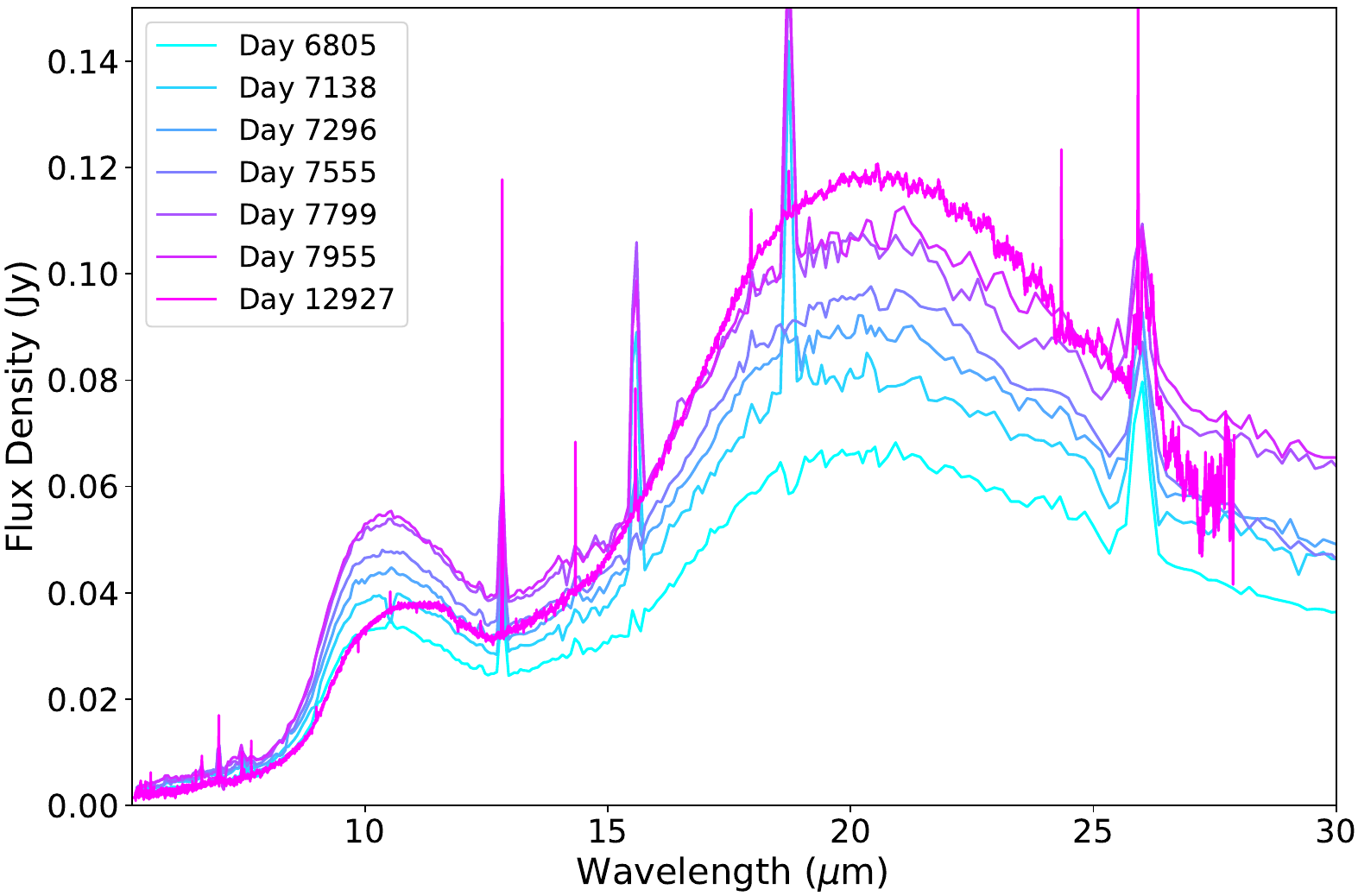}
\includegraphics[width=0.49\hsize]{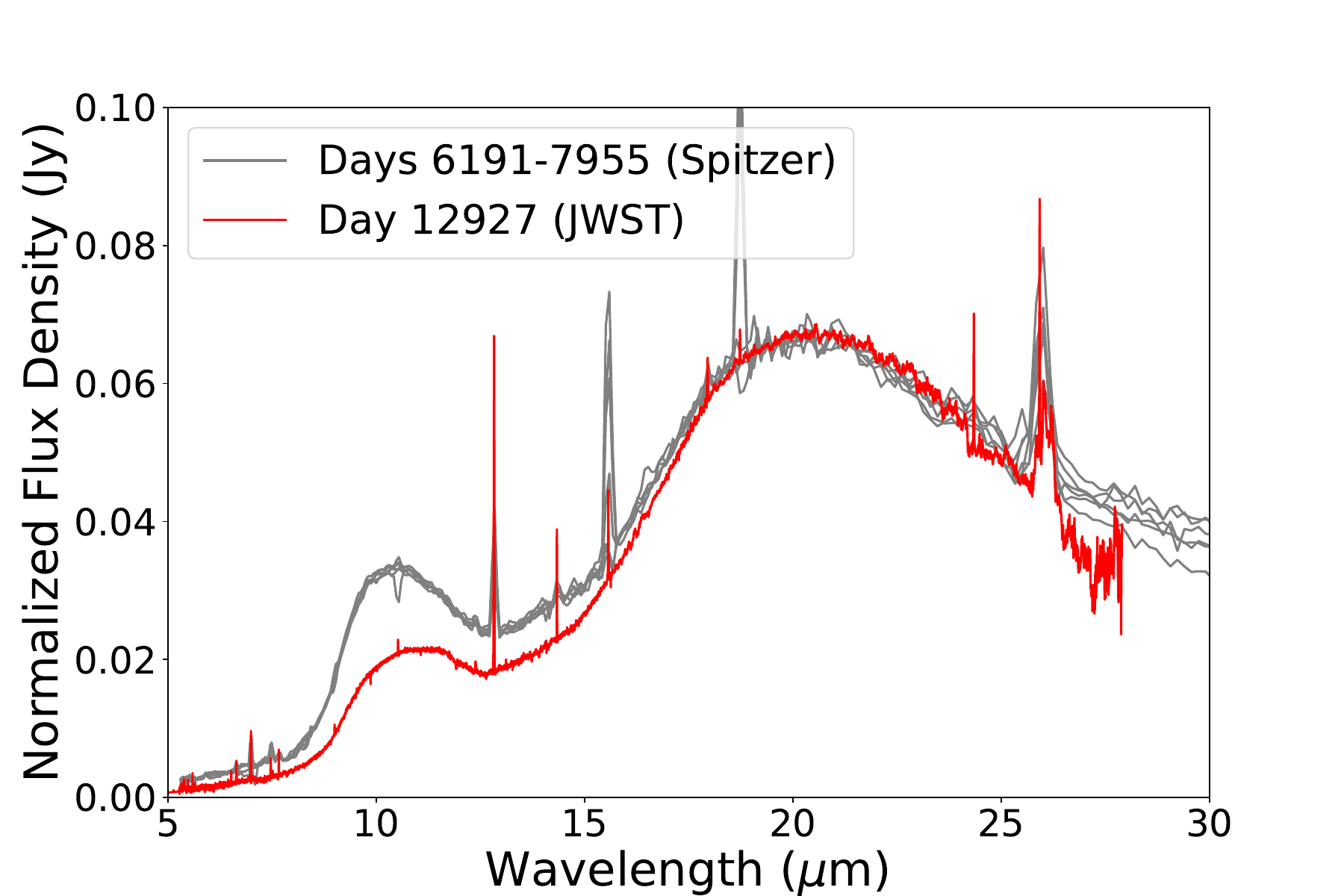}

\caption{Left: Temporal evolution of the ER's mid-IR spectra observed by \textit{Spitzer} IRS (days 6805--7955; years 18.6--21.8) and JWST's MRS instrument (day 12927; year 35.4). Right: Spectra from the left panel normalized to the day 6805 (year 18.6) flux density in the 20--21 $\micron$ range.} 
\label{Fig:IR_spectra_compare}
\end{figure*}

\subsubsection{Temporal Evolution}

The comparison between the {\em Spitzer}/IRS spectra from previous epochs to the day 12,927 (year 35.4) total MIRI/MRS spectra extracted from SN 1987A is shown in the left panel of Figure~\ref{Fig:IR_spectra_compare}. The {\em Spitzer} IRS data from days 6000 to 8000 (years 16.4 to 21.9) have been reduced in a consistent manner using the optimal extraction from the Combined Atlas of Sources with Spitzer IRS Spectra (CASSIS; \citealt{Lebouteiller2011}), with the Short-Low scaled to match the Long-Low flux following \cite{Dwek2010}. 
The MRS spectrum is obtained from within an elliptical region just inside the background boundaries (see Fig.\ref{Fig:spec_extraction_Regions}) and  includes all emission from the ER and inner region and is thus comparable to the {\em Spitzer} IRS data in which the SN 1987A system is not spatially resolved. It can be seen that since day $\sim$8000 (year 21.9), the brightness of the broad 20~$\micron$ silicate feature has continued to increase whereas the part below 15~$\micron$ has decreased. This effect is clearly illustrated in the right panel of Figure~\ref{Fig:IR_spectra_compare} where the spectra have been normalized to the flux density of the 20--21~$\micron$ range of the day 6805 (year 18.6) spectrum. As was shown by  \cite{Dwek2010}, the {\em Spitzer} spectra (grey curves) have an almost identical spectral shape. However, while the day 12927 (year 35.4) MRS spectrum shows the same approximate shape above 18~\micron\ (the spectral region that samples the cold dust component), the relative flux densities at shorter wavelengths are significantly lower.  

\begin{figure*}
\centering
\includegraphics[width=0.49\hsize]{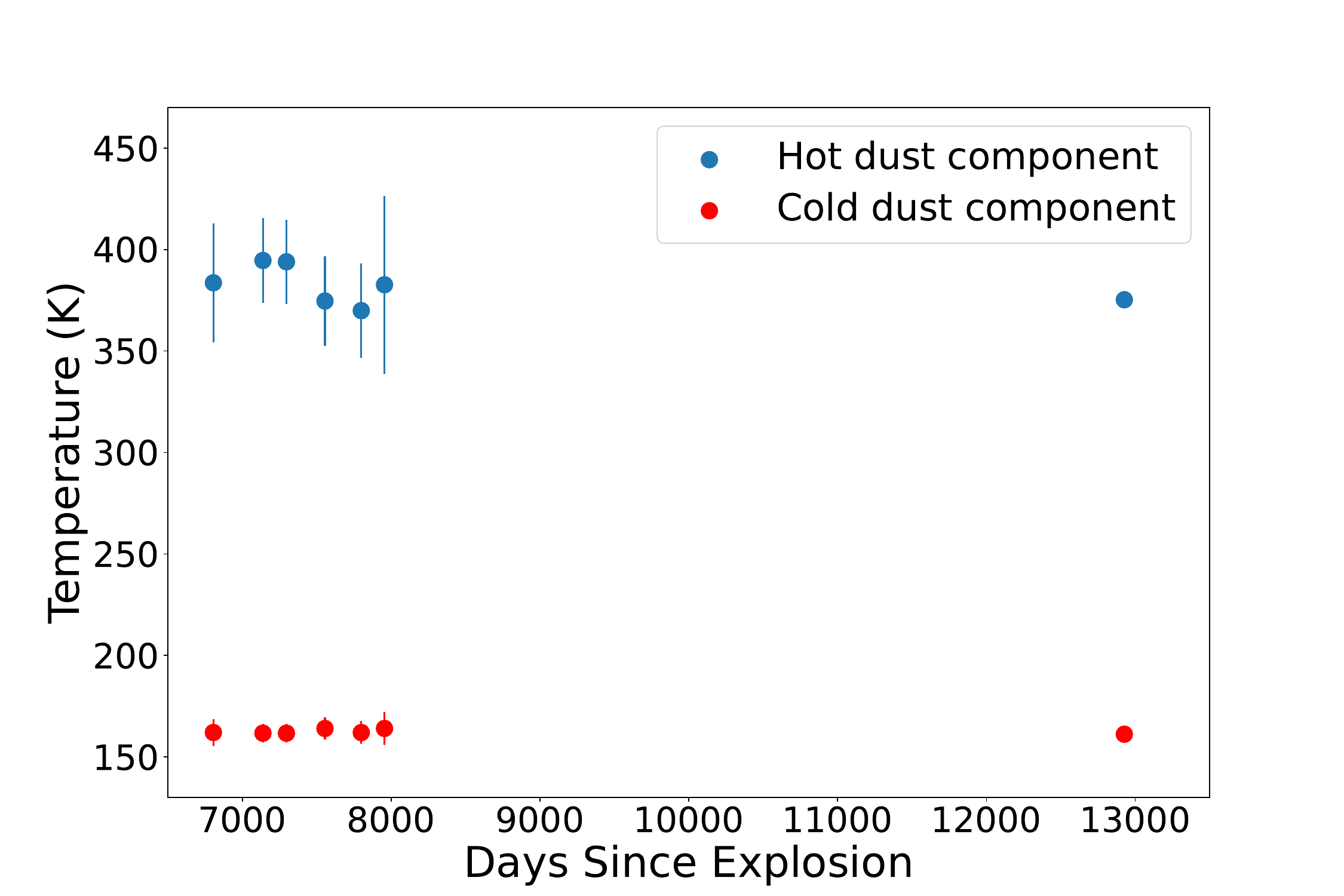}
\includegraphics[width=0.5\hsize]{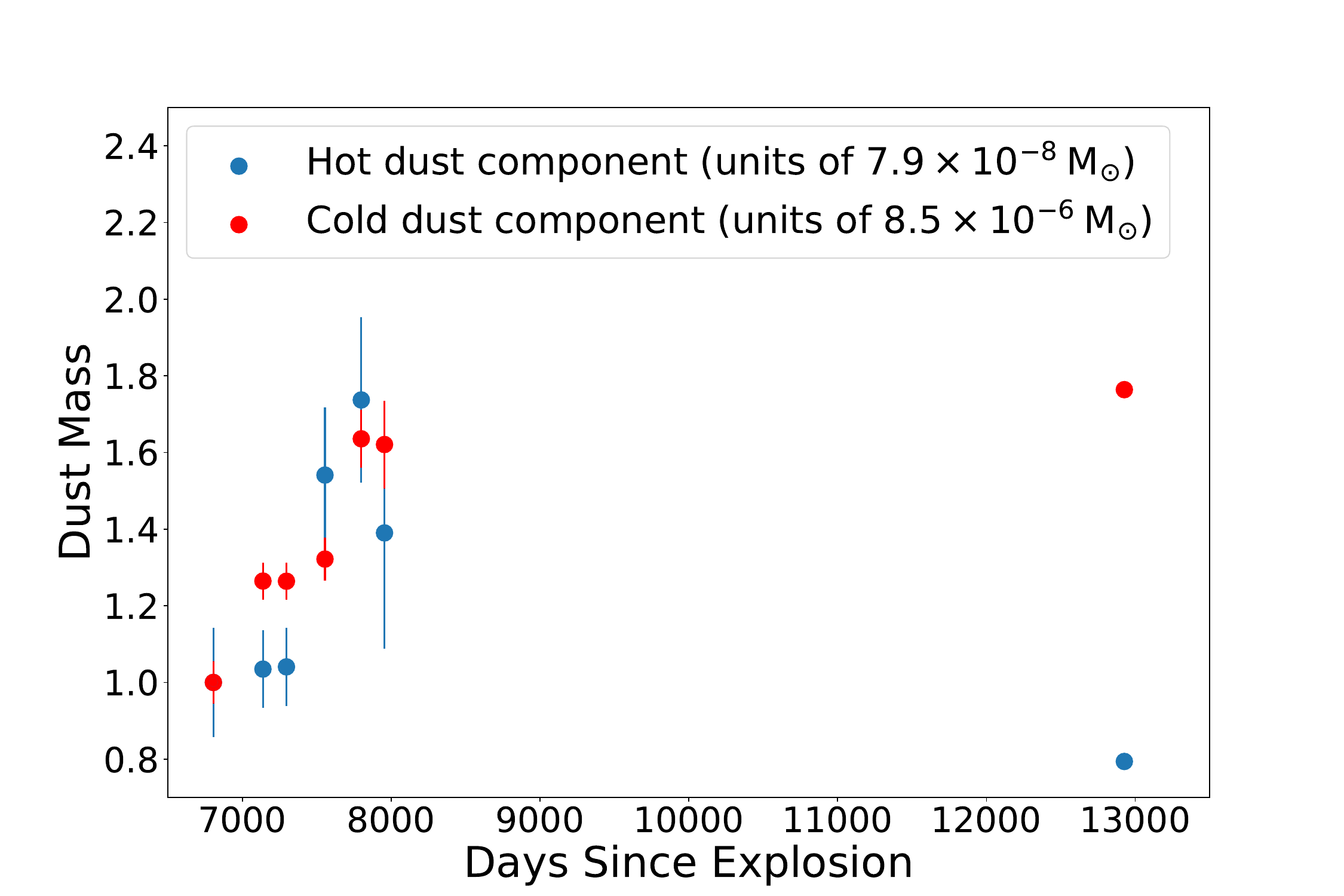}
\caption{Best fit temperatures (left panel) and dust masses (right panel) using a two-temperature dust model for each spectrum shown in Figure~\ref{Fig:IR_spectra_compare}. The spectra were fitted over the 5-30 $\micron$ spectral range using the \textit{astrodust} \citep{Hensley2023} grain composition. The error bars represent the 3$\sigma$ uncertainties from the fit. The data are summarized in Table~\ref{Table:tempdust}.} 
\label{Fig:temp_mass_evol}
\end{figure*}

In order to determine the evolution of the dust parameters over time, we fitted all spectra shown in Figure~\ref{Fig:astrodust_fit} with a two-temperature dust model using the \textit{astrodust} composition. For a consistent comparison between \textit{Spitzer} IRS and MRS, we only fitted the 5--28~$\micron$\ range common to both instruments. We included the previously discussed time-varying synchrotron component \citep{Cendes2018} in the model. The resulting dust temperatures and masses are summarized in Table~\ref{Table:tempdust} and shown in Figure~\ref{Fig:temp_mass_evol}. Interestingly, the dust temperature has remained the same within the uncertainties across all epochs. The cold dust mass measured from the most recent MRS data has increased slightly since day $\sim$8000 (year 21.9), but the mass of the hot dust component dropped by more than 40\% (right panel of Figure~\ref{Fig:temp_mass_evol}). This is likely indicative of significant destruction of the small grain component over time, as previously noted by \cite{Arendt2020} based on \textit{Spitzer} IRAC imaging observations.

\begin{figure*}
\centering
\includegraphics[width=0.49\hsize]{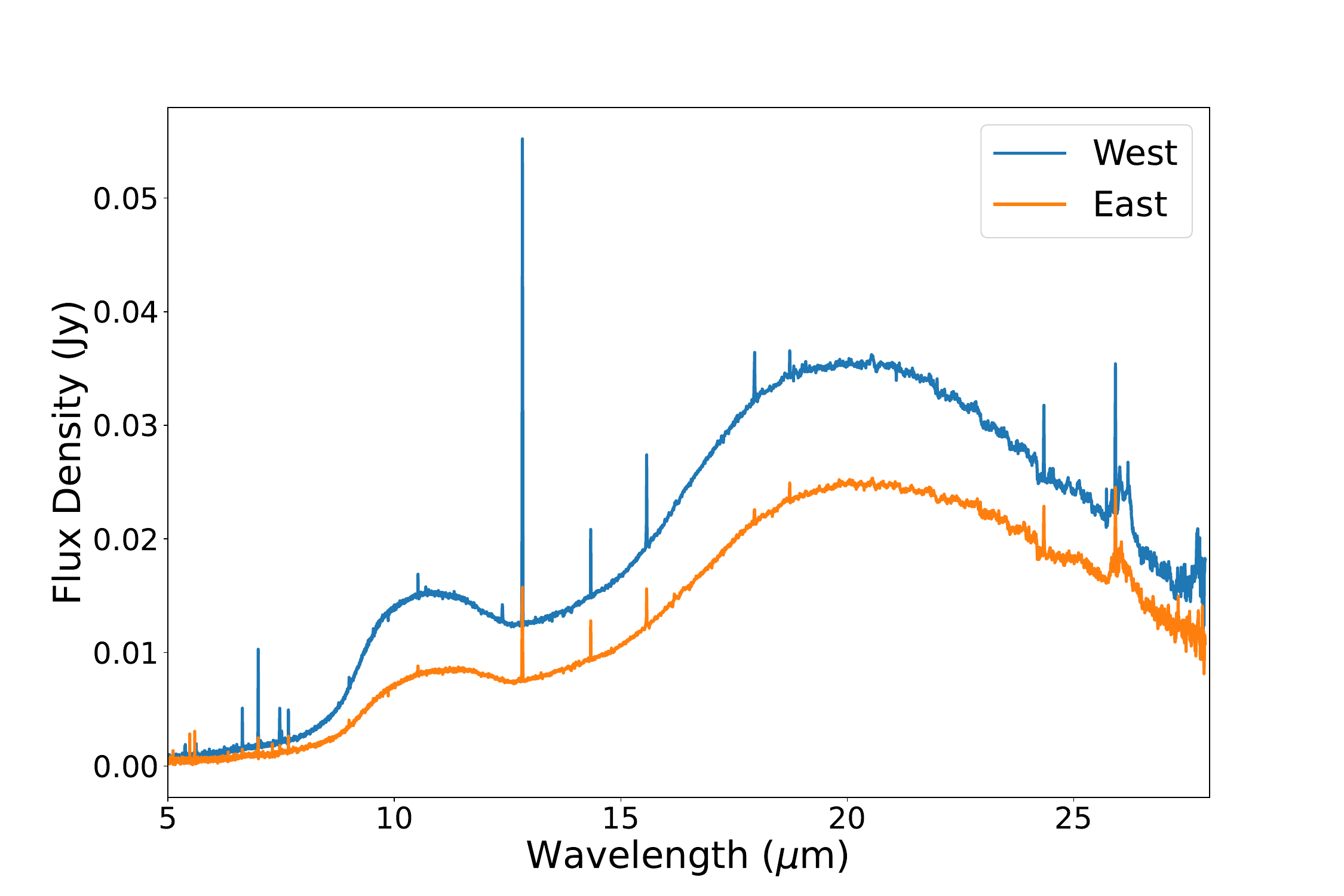}
\includegraphics[width=0.49\hsize]{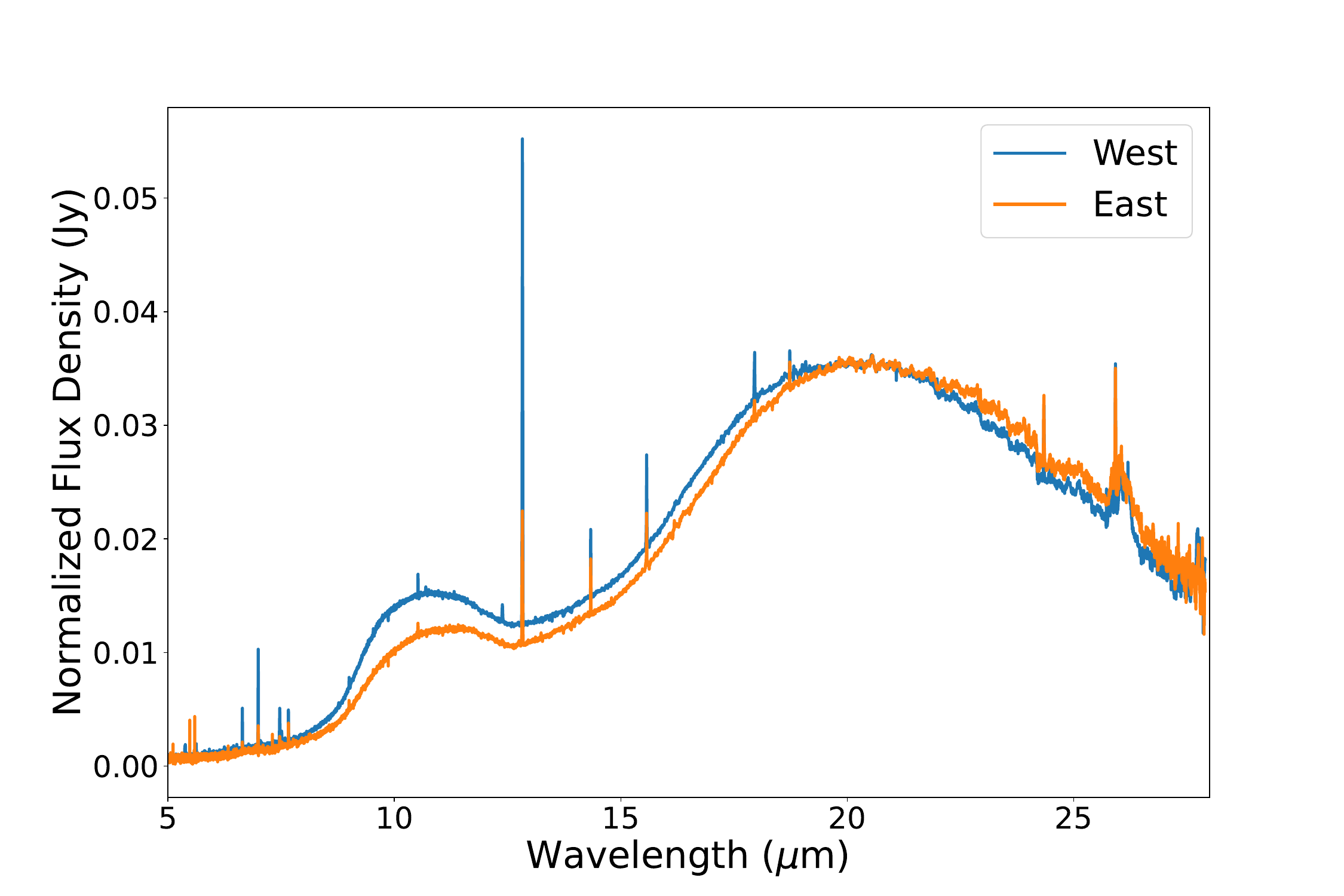}
\caption{Left: A comparison of the ER spectra extracted from the East and West regions shown in Figure~\ref{Fig:spec_extraction_Regions}. Right: A comparison of the East and West spectra that have been normalized to the integrated 20--21~\micron\ flux density of the West spectrum. The blue-ward shift of the broad 20~\micron\ feature suggests that the dust temperature is higher in the West.} 
\label{Fig:cp_spectra}
\end{figure*}

The temporal evolution does not occur in a uniform way over the entire ER. As was shown with IRAC imaging at 3.6 and 4.5~\micron\ \citep{Arendt2020} and with mid-IR imaging \citep{Matsuura2022} the brightening of the ER started in the east in the 2000's after which it faded again and the brightening started in the west by the late 2010's. Figure~\ref{Fig:cube_det_feat} confirms that the W is still brighter than other
portions of the ring over the whole wavelength range of the MRS. A comparison of the spectra taken from the E and W segments of the ER (Figure~\ref{Fig:spec_extraction_Regions}) is shown in Figure~\ref{Fig:cp_spectra}. The W is approximately 50\% brighter than the E part. 
Apart from the lower overall flux in the E, there are also some spectral changes that can be seen in the spectra normalised at 20~\micron\ in the right panel of this figure. 
The eastern part has a relatively stronger long wavelength part and the 10~\micron\ silicate band is weaker and its peak has slightly shifted to the red in comparison to the 
W. This spectral difference can be explained by a relatively higher dust temperature in the W. To model the dust parameters like the dust mass and temperature of the two segments 
requires a careful determination of the local underlying synchrotron continuum for the two ring segments and is planned for a future paper.

\subsection{Other Dust Emission}
In contrast to the inner ER, no dust emission was detected in the region corresponding to the two outer rings which are three times more extended than the ER and currently remain unaffected by the blast wave. 

The ejecta dust was too faint to be detected by {\em Spitzer}. Dust emission may be present in the {\em JWST}/MRS ejecta spectra, however, this requires more mature calibration and extraction procedures to separate any potential contamination from the ring, as such the ejecta's solid-state component is subject to a future paper after follow-up Cycle 2 {\em JWST}/MRS observations have been obtained (PID \#2763). 

\subsection{Emission from the ejecta and ejecta - ER interaction}
\label{sec:ejecta}
During our search for small-scale emission line structures in the SN~1987A system we found that, in addition to the primary emission line peaks with radial velocities consistent with an origin in the ER, outer rings, or diffuse ISM, some lines had additional high-velocity blue- and/or red-shifted features in the northeast and southwest of the ER. Two of these, [Fe~{\sc ii}] 5.340169~$\mu$m and [Fe~{\sc ii}] 25.98839~$\mu$m have already been described in Sect.~\ref{sect:emission_line_props}. Further to these, high-velocity components were detected for [Ni~{\sc ii}]~6.636~$\mu$m, [Ar~{\sc ii}]~6.985~$\mu$m, and [Ne~{\sc ii}]~12.814~$\mu$m. Because the spectral and spatial information complements each other nicely, we discuss first the integrated line spectra and later the spatial maps. One may note that especially at the longer wavelengths the velocity resolution is considerably better than the spatial, although smoothed over a surface of constant line of sight velocity. 

From observations of the  {\em HST} light curve and imaging, we know that the ejecta is powered by radioactive decay of ${}^{44}$Ti and X-rays from the ejecta -- ER interaction \citep{Larsson2011}. The X-ray-shielded inner ejecta is mainly powered by radioactive decay of ${}^{44}$Ti, while the outer parts are heated by the photoelectric absorption of X-rays \citep{Fransson2013}. In the spectra, the ejecta are seen as broad lines expanding with velocity $2000-10,000$ \kms, where the dense, metal-rich core of the SN has a velocity of $2000-3000$ \kms, mainly emitting in lines from  Fe~{\sc i-ii},  O~{\sc i},  Mg~{\sc i-ii},  Si~{\sc i} and  Ca~{\sc ii}, but also from  H~{\sc i}, and He ~{\sc i}, mixed into the core \citep{Chugai1997,Kozma1998,Jerkstrand2011}. 
The ${}^{44}$Ti input is partly in gamma-rays and partly positrons, where the former mainly escapes at these late epochs while the positrons are deposited locally in the Fe and O-rich gas. Because of the very low temperature of the gas in the core, $\lesssim 200$ K,  it is expected that a major fraction of this energy will be emitted in the [Fe~{\sc ii}] 25.99 $\mu$m line \citep{Kozma1998,Jerkstrand2011}. This and other  [Fe~{\sc ii}]  lines are therefore of special interest for understanding the conditions in the core of the SN. 

\begin{figure}
\centering
\includegraphics[width=1.0\hsize]{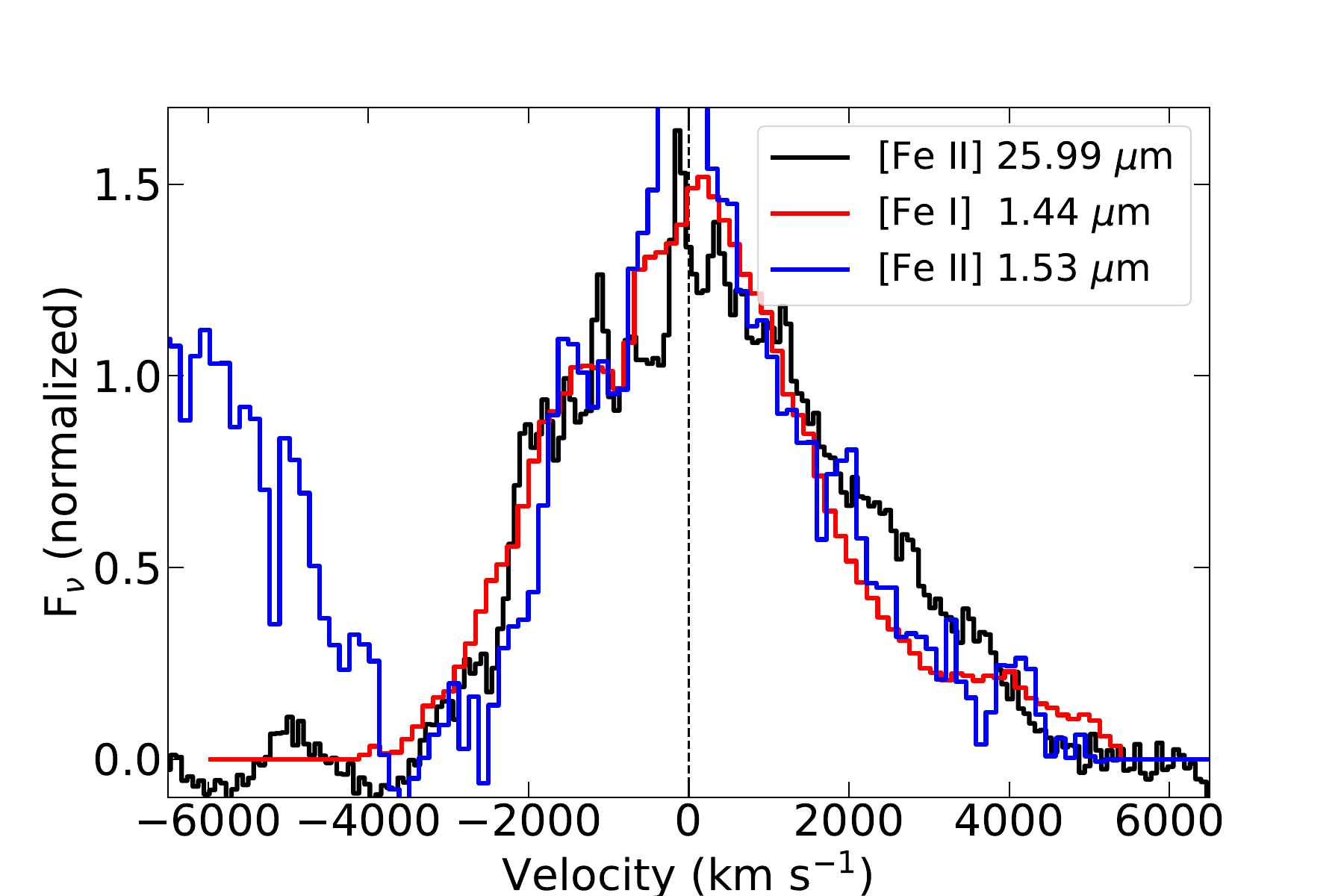}
\caption{Comparison of the continuum subtracted line profiles of the [Fe~{\sc ii}] 25.99 $\mu$m line with [Fe~{\sc i}] 1.44 $\mu$m and [Fe~{\sc ii}] 1.53 $\mu$m lines from the inner ejecta. } 
\label{Fig:fe_ii_26mu_nir}
\end{figure}

In Fig. \ref{Fig:fe_ii_26mu_nir} we show an extraction of the central part (see Fig. \ref{Fig:spec_extraction_Regions}) of the ejecta in the  [Fe~{\sc ii}] 25.99 $\mu$m line, together with the NIR  [Fe~{\sc i}] 1.443 $\mu$m and  [Fe~{\sc ii}] 1.534 $\mu$m lines from the \textit{JWST} NIRSpec spectra \citep{Larsson2023}. The three lines have similar profiles, in spite of different ionization stages and excitation energies. The [Fe~{\sc ii}] 25.99 $\mu$m line, which has the best spectral resolution,  extends from $\sim 3400$ \kms\ on the blue side to $\sim 5300$ \kms\ on the red side. The line is quite asymmetric, with a steep blue side and a more extended red side. 

\begin{figure*}
\centering
\includegraphics[width=0.3\hsize]{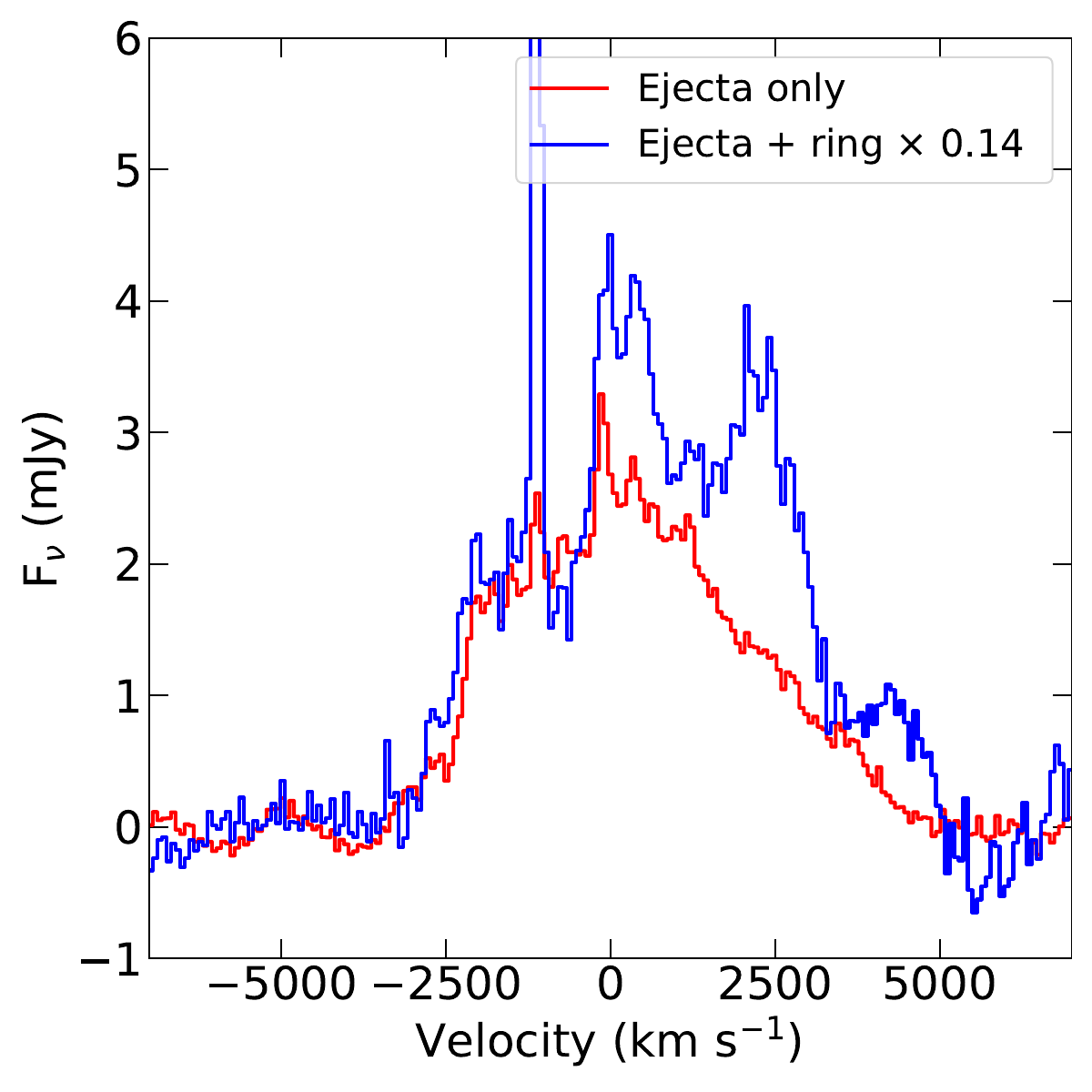}
\includegraphics[width=0.3\hsize]{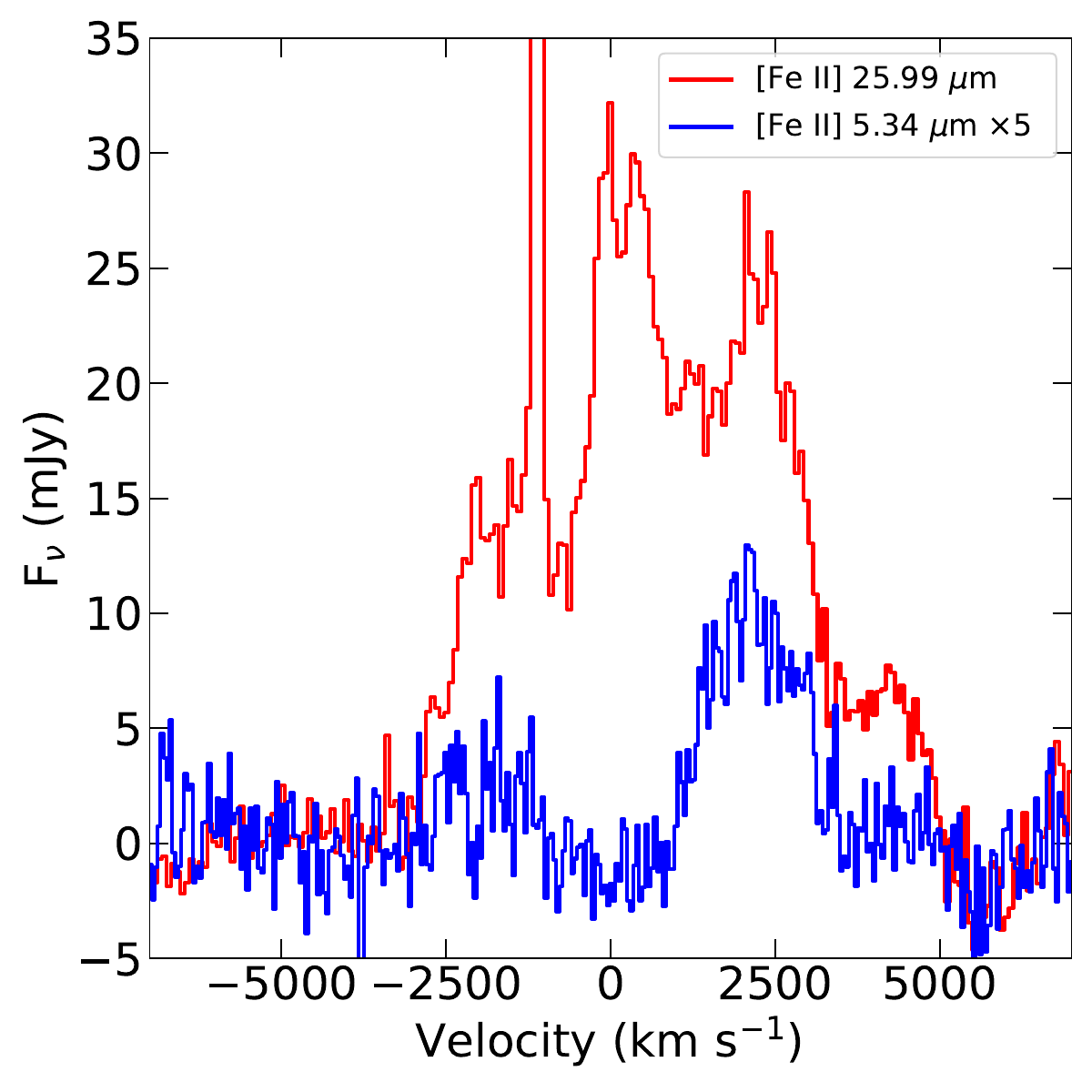}
\includegraphics[width=0.3\hsize]{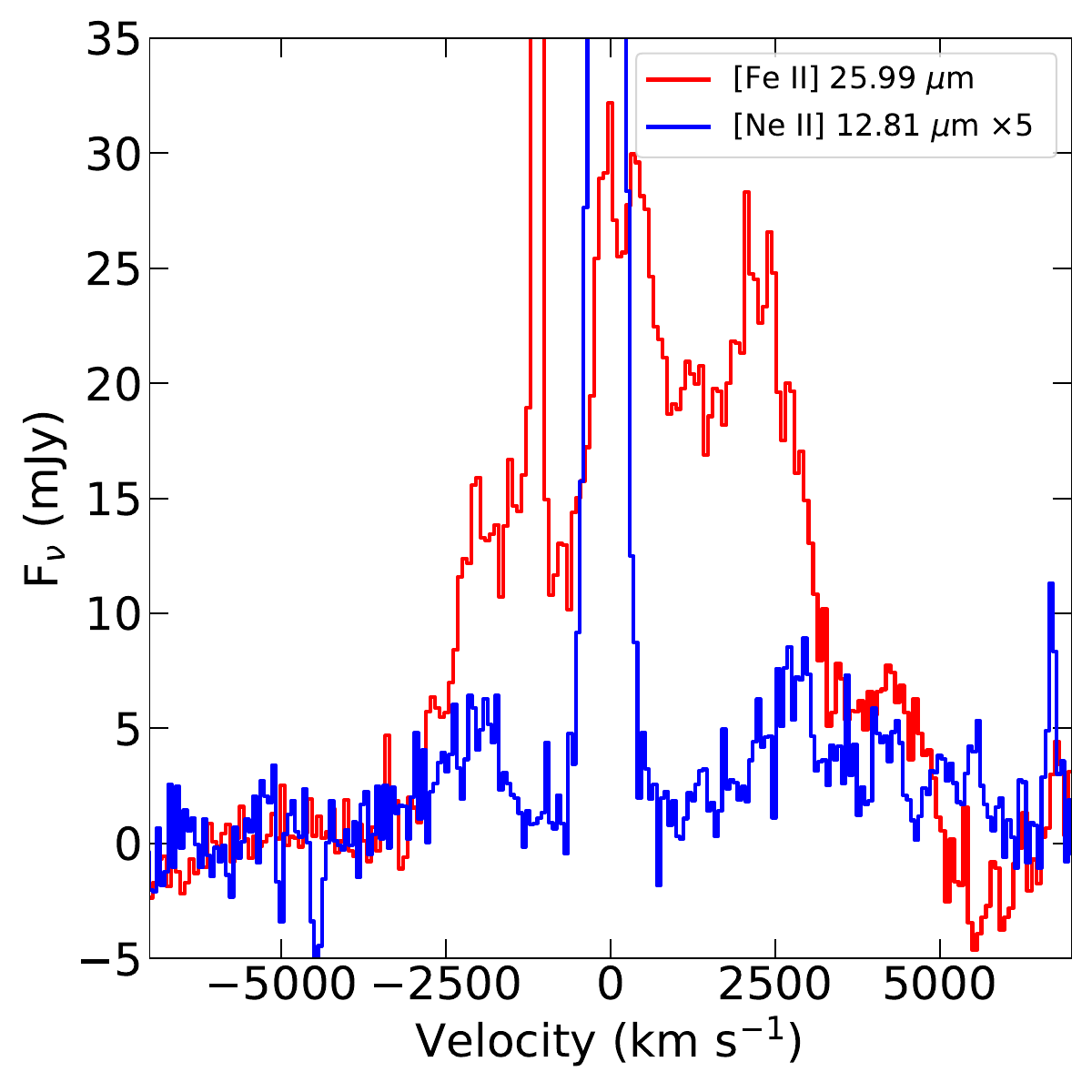}
\caption{Left: Comparison of the [Fe~{\sc ii}] 25.99 $\mu$m line profiles from the inner ejecta and the total line profile including the ring. The spike at -1150 \kms{} is the narrow [O~{\sc iv}] 25.91 $\mu$m line from the ER. Middle: Comparison of the line profiles of the [Fe~{\sc ii}] 5.340 $\mu$m and [Fe~{\sc ii}] 25.99 $\mu$m lines. Note the lack of emission between $\pm 1500$ km s$^{-1}$ in the 5.34 $\mu$m profile. Right: The same for the [Ne~{\sc ii}] 12.81 $\mu$m line.} 
\label{Fig:fe_ii26_534_neii_comp}
\end{figure*}

This central extraction showed only faint traces of the other lines mentioned above. However, an extraction also covering the ER (Fig. \ref{Fig:spec_extraction_Regions}) had a clear signal for all of these. In Fig. \ref{Fig:fe_ii26_534_neii_comp} we show on the left panel a comparison of this extraction with the central extraction. In this plot we have scaled the total extraction by a factor of 0.14 to agree at $\sim 1500$ \kms, mainly reflecting the small central extraction region compared to the total extraction (Fig. \ref{Fig:spec_extraction_Regions}).  The most notable difference is the strong peak at $\sim 2200$ \kms, but also a fainter peak at $\sim 4100$ \kms. The peak at $\sim 2200$ \kms\ corresponds to the interaction region in the SW, as can be seen on the line maps below. The central peak, not seen in the central extraction, may come from the part of the dense ejecta above the ER plane, which has been bright also in earlier SINFONI observations \citep{Larsson2016}, when it was still located inside the ER. Because of the expansion, it is now projected against the ER. 

In the middle panel, we show a comparison with the   [Fe~{\sc ii}] 5.340 $\mu$m line. This comparison is interesting because it shows a major difference with the [Fe~{\sc ii}] 25.99 $\mu$m line in that the low-velocity emission is very low compared to the emission peaks at $\sim +2200$ \kms\ and $\sim -1800$ \kms. 
This indicates that there is little [Fe~{\sc ii}] 5.340 $\mu$m emission from the central parts of the ejecta, as was indicated from the central extraction. It also separates out the emission from the ER interaction on the blue side, which is not easily distinguished from the 25.99 $\mu$m line.

The right panel of Fig. \ref{Fig:fe_ii26_534_neii_comp} shows a corresponding comparison with the [Ne~{\sc ii}] 12.81 $\mu$m line. 
The [Ne~{\sc ii}] line exhibits a central component that originates from the ER and which is stronger and narrower than the central [Fe~{\sc ii}] 25.99-$\mu$m line emission. Outside of its central ER emission, the [Ne~{\sc ii}] profile exhibits a similar line shape as the [Fe~{\sc ii}] 5.340 $\mu$m line, with a stronger red component and a weaker blue component at similar velocities.

\begin{figure*}
\centering
\includegraphics[width=0.99\hsize,trim={0cm 0cm 0cm 0cm},clip]{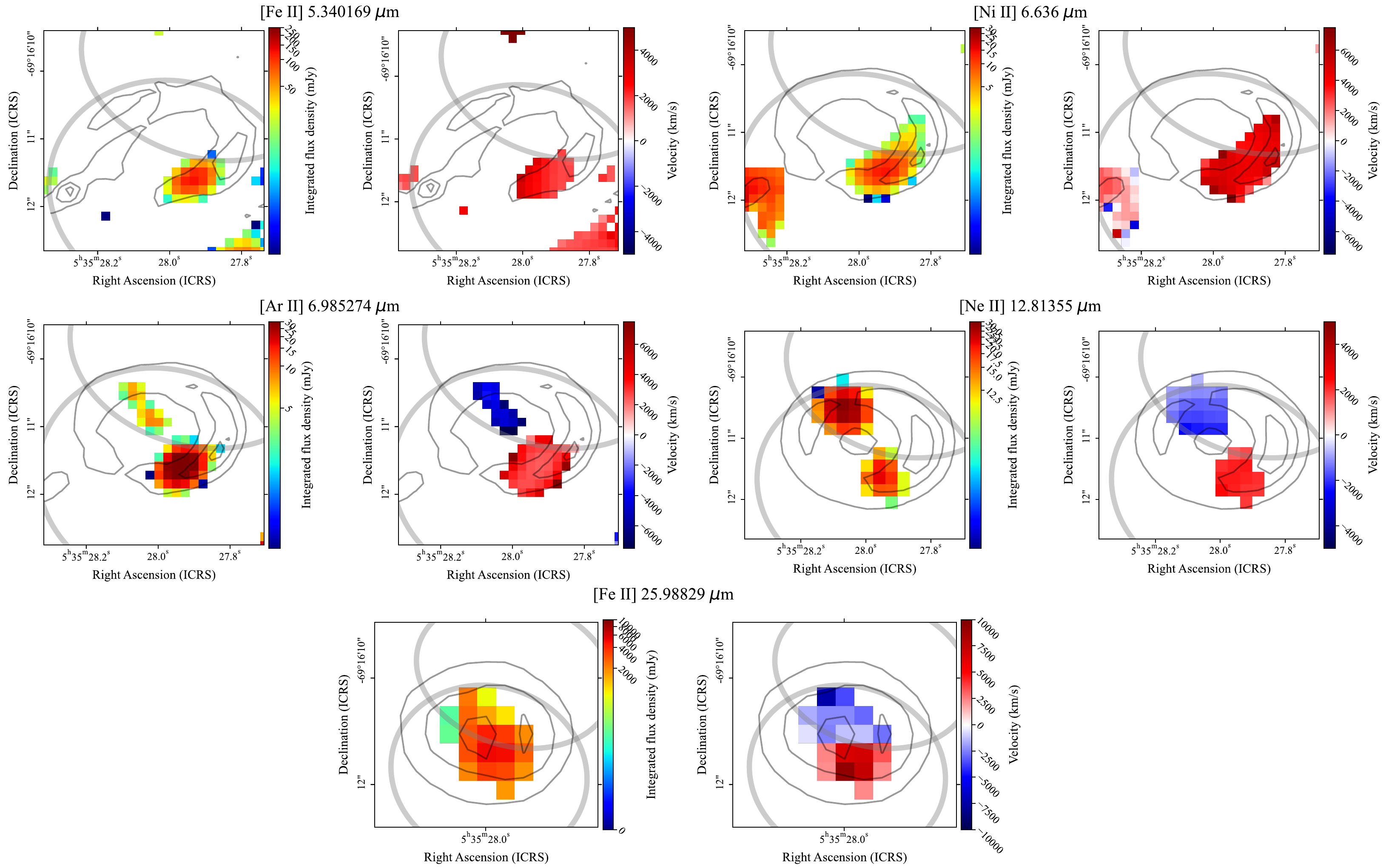}
\caption{Same as Fig.\ref{Fig:fe_ni_ring_mm} but for high-velocity line components and without the FWHM maps.
}.
\label{Fig:ej_ring_int}
\end{figure*}

To determine the morphology and radial velocity of these structures spatially we again produced intensity and velocity maps using the 0th and 1st moments using Spectral-Cube, this time masking the contribution of the low-velocity component. The resulting maps are shown in Fig.~\ref{Fig:ej_ring_int}. Significant, highly red-shifted emission centred at $\sim+2000$~km~s$^{-1}$ and projected towards the southwestern ER is observed in all lines. In the cases of [Ar~{\sc ii}]~6.985~$\mu$m, [Ni~{\sc ii}]~6.636~$\mu$m, and [Fe~{\sc ii}] 25.988~$\mu$m, a blue-shifted component is also observed, though projected towards the northeast of the ER. The location and velocity of these features agree with the \textit{JWST} NIRSpec study of \citet{Larsson2023}, who found broad [Fe~{\sc i}]~1.443~$\mu$m and [Fe~{\sc ii}]~1.644~$\mu$m towards the northeast and southwest of the ER. \citet{Larsson2023} attributed these features to the interaction of the expanding ejecta with the reverse shock and also noted that northeast-southwest elongated morphology has been observed in all other atomic lines from the ejecta \citep{Wang2002, Kjaer2010, Larsson2013, Larsson2016}. We suggest here that the high-velocity emission observed in our MIRI/MRS observations results from that same ejecta-reverse shock interaction. A more detailed analysis of the ejecta and interaction regions will be presented in future work.

\section{Discussion}
\label{sec:discussion}

The MIRI MRS observations presented above reveal the first spatially resolved spectral maps in the 5 to 28 $\mu$m wavelength region. The gas emission lines arise in the ER, the ejecta, the outer rings and the region between the ER and the outer rings. The continuum, which is dominated by dust emission, arises primarily in the ER.

The improved spectral resolution over {\it Spitzer} reveals the kinematic origin of the lines as well as enabling the discovery of new gas emission lines such as [Fe~{\sc ii}] 5.34~$\mu$m, [Na~{\sc iii}] 7.32~$\mu$m,  [Ar~{\sc iii}] 8.99 $\mu$m,  and [Ni~{\sc ii}] 10.68 $\mu$m (Sect.~\ref{sec:lines}).  The origin of the narrow but centrally located [Ar~{\sc ii}] lines is discussed by \cite{Fransson2024}. 
The [Ni~{\sc ii}] lines set an electron temperature of T$_e \leq 6500$~K in the western lobe.  This relatively low temperature plus the low degree of ionization of the broad-line emitting species in the ER are consistent with dense post-shock cooling gas emission.
The current higher density of the western side compared to the rest of the ER has likely also elevated the temperature and heating of the dust in the western lobe \citep{Matsuura2022}.

The origin of the narrow-line component, associated with the spectra of the more highly ionized species, is less clear. On the one hand, the surface brightnesses of the narrow lines peak at the equatorial ring, indicating some degree of association with the material in the ER. On the other hand, the 
narrow-line emission component shows no significant radial velocity variations and extends far beyond the ER, permeating the outer rings and reaching the edges of the MRS field with declining surface brightness. Candidate mechanisms to produce this narrow-line, higher degree of ionization, emission component include photoionization by X-ray and ultraviolet photons from shocks passing through the ER, and flash ionization by the original UV photon pulse of circumstellar or interstellar material of sufficiently low density that recombination has not yet completed.

Previously \cite{Dwek2010} and \cite{Arendt2016} used a two-component dust model to fit the low-resolution {\em Spitzer}/IRS  spectra, consisting of a warm dust amorphous silicate component with a temperature of 180 K to model the 8--35 $\mu$m emission and a secondary featureless hot dust component fit to the 5--8 $\mu$m flux assumed to have an amorphous carbon ($T = 460$ K) composition. This secondary component may also be fit by iron grains as its optical constants are also featureless which would result in hot dust temperatures of $\sim$350 K. \cite{Arendt2016} determined temperatures of 190 K and 525 K for the silicate and carbonaceous components, respectively. 
Using the same sets of mass absorption coefficients as \cite{Dwek2010} and \cite{Arendt2016} we obtain slightly cooler temperatures for both components in our dust model fits to the MRS data, as we account for the contribution from the extrapolated long-wavelength synchrotron component and the H and He continuum emission in the near-IR. Our multi-component model is also representative of the near-IR emission which prior models of the ER emission did not manage to fit successfully.

In our analysis, \textit{astrodust} \citep{Hensley2023} optical constants best fit the NIRSpec and MIRI spectra. \textit{Astrodust} is comprised of C, O, Mg, Si, and Fe in a single-grain material, including amorphous silicates in the form Mg$_{1.3}$(Fe,Ni)$_{0.3}$SiO$_{3.6}$ and no large carbonaceous grains. This eliminates the need for two-grain compositions (silicates plus a featureless component typically assumed to be carbon) spatially coexisting, which form from completely divergent condensation pathways in the RGB phase of evolution. 
Grains formed during this phase are primarily oxygen-rich, due to CNO processes in the core resulting in low carbon abundances, with any remaining C atoms locked up in CO molecules. The additional benefit of our model is that the cooler temperature we obtain for our hot dust component can be reached by collisional heating under thermal equilibrium, eliminating the need for the more problematic stochastic heating which very small amorphous hydrocarbons from previous fits would require \citep{Arendt2016, Dwek2010}.  

While beyond the scope of the present study, detailed modelling of the various components of the SN~1987A MRS spectra with laboratory data is an important step toward understanding the evolution of dust in SNe. Using \cite{Dorschner1995} optical constants for amorphous silicates which \cite{Jones2014} found best fit the spectra of RSGs in the LMC results in fits to the ER which are too sharp to reproduce the 10 and 20-micron features. Likely, some additional components at the few percent level including SiO$_2$, Al$_2$O$_3$, CaCO$_3$, Fe oxides, carbides, and sulphur compounds, which are all included in the artificial {\em astrodust} composite grains, would be required. Pure Fe is ruled out as a viable candidate due to the large dust mass that would be required (Table~\ref{Table:dusttab}).

The evolution of the hot dust $<$15~$\mu$m component is consistent with the steady decline seen in the 3.6 and 4.5 $\mu$m flux measured during {\em Spitzer's} warm mission.
The {\em Spitzer} cryogenic data pre-day 8000 was found to follow the same trends as the X-ray emission prior to diverging post-day 8000 (post-year 21.9) \citep{Arendt2016, Arendt2020}. This corresponds to the collisional heating of the dust in the ER before a transition to radiative heating or a change in the physical conditions in the ring, for instance, grain destruction or a change in the dust grain size distribution. 
Thermal sputtering timescale suggests that grain destruction may become prominent from day 9200 (year 25.2) after the explosion \citep{Dwek2010}.
This dust processing and destruction are conclusively shown for the hot dust in the MRS spectra obtained on day 12927 (year 35.4) , however, the effects are negligible for the cold dust emission which may even be growing in mass (though not significantly) due to the expanding propagation of the forward shock wave sweeping up remnant circumstellar material.  

Dust grains in the ring are expected to have a distribution of grain sizes that are heated to different temperatures, with the resulting emission spectrum reflecting a continuous temperature gradient as a function of grain size. In our fits, this gradient is approximated by a two-temperature-component model. The temperature differential between our hot and cold components in the {\em astrodust} model is, therefore, likely a consequence of the dust grain size distribution, with the hot dust component corresponding to smaller grains and the cold component to the larger ones. The trend seen in Figure~\ref{Fig:temp_mass_evol} can thus be interpreted as circumstellar material from the progenitor being swept up by the blast wave expanding through the ring. As the shock propagates through the ring, the mass of the shock-heated dust grows in proportion with the total swept-up gas mass, as seen on days 7000-8000 \citep{Dwek2010}. As they are replenished, the smaller grains are being simultaneously destroyed in the ER on timescales of a few years \citep{Arendt2016}. Once the grains stop being replenished in sufficient amounts, the destruction of the small grains dominates, whereas the large grains are more likely to survive the passage of the high-velocity shock, resulting in the divergence in dust masses between the hot and cold components seen in the MRS data.
This scenario requires the hot grains to be non-carbonaceous, as the binding energy makes carbonaceous grains relatively stable against thermal and non-thermal sputtering \citep{Fischera2002}. 

SOFIA observations \citep{Matsuura2019} identify a 31.5--70 $\mu$m excess which may be attributed to dust re-formation after the passage of the forward shock, warm dust emission from the ejecta or the survival of large grains.  \cite{Matsuura2019} prefer the reformation of dust grains after the shock front. There is however no indication in the MRS data of a warm dust component from the ejecta nor a re-formed dust component or substantial grain growth in the ring, given the consistency between the long-wavelength MIRI/MRS and {\em Spitzer}/IRS data. This suggests that dust reformation in the post-shock region of this system does not easily occur or occurs on much longer time scales given the low densities \citep{Biscaro2014}. Our results support the survival of large grains in the shocks.

The different line shapes between the [Fe~{\sc ii}] 25.99 $\mu$m  line on the one hand and the [Fe~{\sc ii}] 5.340 $\mu$m and [Ne~{\sc ii}] 12.81 $\mu$m lines, on the other hand, show that the excitation of the outer parts of the ejecta, close to the ER, and the central region are different. The excitation energy of the upper level of the  [Fe~{\sc ii}] 25.99 $\mu$m  line corresponds to 550 K, while it is 2676 K for the [Fe~{\sc ii}] 5.340 $\mu$m line, 1115 K for the  [Ne~{\sc ii}] 12.81 $\mu$m line and 2045 K for the  [Ar~{\sc ii}] 6.985 $\mu$m line. From the modelling of the 8-year spectrum in \cite{Jerkstrand2011} the temperature in the ${}^{44}$ Ti-powered inner core is $70-170$ K. The excitation of the [Fe~{\sc ii}] 25.99 $\mu$m  line can therefore occur at low temperatures, while the much higher excitation energies of the other lines require a higher electron temperature. This is consistent with the observed dip in the central line profile of the [Fe~{\sc ii}] 5.340 $\mu$m line.  The temperature at 35 years may be different from the above, but these line profiles indicate a very low temperature in the core also at the present epoch. Models of the X-ray transfer in the ejecta from the ejecta -- ER interaction also show that most of the X-rays have not yet penetrated to the central parts of the core \citep{Fransson2013,Kangas2022}. The high-velocity  Fe-rich regions which have expanded to close to the reverse shock are, however, not shielded from the X-rays due to a much smaller column density between the reverse shock and ejecta. They will therefore be heated efficiently, as is directly seen from the line profiles and the maps, and will have a much higher temperature. More detailed modelling of the shock interaction is required for a more quantitative estimate.

\section{Conclusions}
\label{sec:conclusions}

We have presented mid-IR spectra of SN 1987A obtained with the MRS instrument onboard {\em JWST} on day 12,927 post-explosion. The 5--28 $\mu$m MRS spectroscopic observations spatially resolve the ER, ejecta, outer rings and surrounding environment. We find:

\begin{enumerate}
\item Infrared line emission from the ER coincides with the dense clumps of gas observed as hotspots in the optical images, while we find that the dust continuum emission is more spatially extended. 

\item Emission line radial velocities, FWHM values and integrated line fluxes were measured at four cardinal points in the ER and for the whole ring. Nineteen emission lines, from  thirteen species, were detected. The majority of the lines identified have previously been detected in {\em Spitzer} spectra of SN 1987A, including [Ni~{\sc ii}] 6.64~$\mu$m, H~{\sc i} 7.46~$\mu$m and [Ne~{\sc vi}] 7.65~$\mu$m which were previously only tentatively identified. However, [Fe~{\sc ii}] 5.34~$\mu$m, [Na~{\sc iii}] 7.32~$\mu$m, [Ar~{\sc iii}] 8.99~$\mu$m and [Ni~{\sc ii}] 10.68~$\mu$m are new detections.

\item The MRS lines are approximately an order of magnitude fainter compared to the {\em Spitzer} detections on day 7954 (year 21.8), which could be partly attributable to a smaller extraction region for the MRS lines. However, relative to [S~{\sc iv}] 10.51 $\mu$m, the low degree of ionization [Fe~{\sc ii}] and [Ne~{\sc ii}] lines have brightened by factors of 5--12 in the MRS spectra.

\item The velocity structures of the ER lines indicate that they arise from two regions: The singly ionized species show broad lines (280-380~km~s$^{-1}$ FWHM) and their radial velocities indicate that they originate from the expanding ER. Their low degree of ionization, together with the upper limit of 6500~K on the ER electron temperature provided by the ratio of the [Ni~{\sc ii}] 6.64- and 10.68-$\mu$m lines, are consistent with recombination of post-shock gas. Lines from the more highly ionized species 
are narrower (99-171 km~s$^{-1}$) and their lack of any radial velocity variations across the ER and beyond indicates that they arise from a different emission component. The ratio of the [Ne~{\sc v}] 14.32- and 24.32-$\mu$m lines  at the ER implies electron densities for this component of between 700--4300~cm$^{-3}$ for T$_e = 5000-25000$~K, while a [Ne~{\sc v}] 14.32-$\mu$m continuum-subtracted image shows the emission to also be associated with material located in the outer rings and beyond. The higher degree of ionization of these species may have been produced by either X-ray and ultraviolet photons from shocks progressing through the ER, or else by flash ionization by the UV photon pulse associated with the original supernova explosion.

\item The overall shape and positions of the features in the ER spectrum are generally consistent with the {\em Spitzer}/IRS spectra obtained between days 6000 and 8000. No resonances from other solid-state features are apparent in the higher-resolution MRS spectra. However, the relative flux densities $<15~\mu$m are significantly lower in the MRS spectra. 

\item The 0.9--28~$\micron$ NIRSpec and MIRI spectral range is best-reproduced by a single set of \textit{astrodust} optical constants emitting at 157$\pm$4 and 334$\pm$12~K. This is contrary to previous models of dust emission from the ER which require a secondary population of very small, carbon or iron grains of unknown origin radiating at significantly higher temperatures to spatially coexist with the cool astronomical silicate grains. 
The inclusion of the continuum from synchrotron and bound-free emission in our models significantly lowers the temperature of the `hot' component even when this secondary population is represented by amorphous carbon or iron grains. 

\item Fitting both the MRS and {\em Spitzer} spectra with the same two-temperature model produces a consistent dust temperature in the ring between day $\sim$~7000 and 12927 (year 19.2 and 35.4). However, there is a significant difference in the mass of the hot dust at day 12927 compared to 7955, suggesting the hot, smaller dust grains are being destroyed.

\item The large dust grains from the progenitor preferentially survive the SN explosion and processing by the forward shock as it evolves into an SNR.

\item There is a shift in the amorphous silicate feature position between the East and West ER components indicating a difference in the shock processing.

\item Our spectra show a number of lines from the ejecta which, with the exception of the [Fe~{\sc ii}] 25.99-$\mu$m line, only have been seen during the first years after explosion. This includes [Fe~{\sc ii}] 5.34~$\mu$m, [Ar~{\sc ii}] ~ 6.99$~\mu$m and [Ne~{\sc ii}] ~12.81~$\mu$m. The [Fe~{\sc ii}] ~25.99~$\mu$m line is in the centre still powered by the radioactive ${}^{44}$Ti decay, while the emission at $\sim 2200$ \kms~ is powered by the ER - ejecta interaction. This is also consistent with the line profiles and spatial distribution of the other lines above. The central [Fe~{\sc ii}] 5.34 $\mu$m to 25.99~$\mu$m line ratio is consistent with a very cool central ejecta, shielded from the X-rays.

\end{enumerate}





\begin{acknowledgments}
This work is based on observations made with the NASA/ESA/CSA James Webb Space Telescope. The data were obtained from the Mikulski Archive for Space Telescopes at the Space Telescope Science Institute, which is operated by the Association of Universities for Research in Astronomy, Inc., under NASA contract NAS 5-03127 for JWST. These observations are associated with program \#1232. This research made use of Photutils, an Astropy package for
detection and photometry of astronomical sources (Bradley et al. 2022).

OCJ acknowledges support from an STFC Webb fellowship. 
MJB acknowledges support from European Research Council Advanced Grant 694520
SNDUST.
PJK and JJ acknowledge support from the Science Foundation Ireland/Irish Research Council Pathway programme under Grant Number 21/PATH-S/9360.
JL acknowledges support from the Knut \& Alice Wallenberg Foundation. 
JL and CF acknowledge support from the Swedish National Space Agency. 
MM and NH acknowledge support through a NASA/JWST grant 80NSSC22K0025, and MM and LL acknowledge support from the NSF through grant 2054178. MM and NH acknowledge that a portion of their research was carried out at the Jet Propulsion Laboratory, California Institute of Technology, under a contract with the National Aeronautics and Space Administration (80NM0018D0004).
JH was supported by a VILLUM FONDEN Investigator grant (project number 16599).
TT acknowledges financial support from the UK Science and Technology Facilities Council and the UK Space Agency.
RW acknowledges support from STFC Consolidated grant (2422911).
LC acknowledges support by grant PIB2021-127718NB-100 from the Spanish Ministry of Science and Innovation/State Agency of Research MCIN/AEI/10.13039/501100011033. TPR acknowledges support through the European Research Council (ERC) under advanced grant no.\,743029 Ejection Accretion Structures in YSOs (EASY). 

\end{acknowledgments}


\vspace{5mm}
\facilities{JWST (MIRI)
All of the MIRI-MRS data presented in this paper were obtained from the
Mikulski Archive for Space Telescopes (MAST) at the Space Telescope Science 
Institute. The specific observations analyzed can be accessed via
\dataset[DOI:10.17909/pz4v-ej33]{https://doi.org/10.17909/pz4v-ej33
[doi.org]}.
}


\software{Astropy \citep{Astropy2018, Astropy2022},  
          specutils \citep{specutils},
          Photutils \citep{Bradley2022},
          Regions \citep{Bradley2022b}.
          }


\bibliography{sn87a.bib}{}
\bibliographystyle{aasjournal}

\end{document}